\documentclass[journal]{IEEEtran}
\usepackage{cite}
\usepackage{tabularx}
\usepackage{graphicx}
\usepackage{subcaption}
\usepackage{mdwmath}
\usepackage{amsmath,amssymb,amsfonts}
\usepackage{mathtools}
\usepackage{epstopdf}
\usepackage{tabularx}
\usepackage{array}
\usepackage{authblk}
\usepackage{algorithm}
\usepackage{algpseudocode}
\usepackage[hidelinks]{hyperref}

\makeatletter
\newcommand{\thickhline}{%
	\noalign {\ifnum 0=`}\fi \hrule height 1pt
	\futurelet \reserved@a \@xhline
}

\newtheorem{definition}{Definition}

\newtheorem{remark}{Remark}

\begin{document}
		
		\title{A Fully Multivariate Multifractal Detrended Fluctuation Analysis Method for Fault Diagnosis}
		
		\author{Khuram Naveed and Naveed ur Rehman
			
			\thanks{K. Naveed is associated with the Department of Dentistry and Oral Health, Aarhus University Denmark, and Department of Computer Engineering, COMSATS University Islamabad (CUI), Islamabad Pakistan (emails: knaveed@dent.au.dk; khurram.naveed@comsats.edu.pk).} 
			\thanks{N. ur Rehman is associated with the Department of Electrical and Computer Engineering, Aarhus University, Aarhus Denmark (email: naveed.rehman@ece.au.dk)}
		}
		
		\maketitle
	
\begin{abstract}
We propose a fully multivariate generalization of multifractal detrended fluctuation analysis (MFDFA) and leverage it to develop a fault diagnosis framework for multichannel machine vibration data. We introduce a novel covariance-weighted $L_{pq}$ matrix norm based on Mahalanobis distance to define a fully multivariate fluctuation function that uniquely captures cross-channel dependencies and variance biases in multichannel vibration data. This formulation, termed FM-MFDFA, allows for a more accurate characterization of the multiscale structure of multivariate signals. To enhance feature relevance, the proposed framework integrates multivariate variational mode decomposition (MVMD) to isolate fault-relevant components before applying FM-MFDFA. Results on wind turbine gearbox data demonstrate that the proposed method outperforms conventional MFDFA approaches by effectively distinguishing between healthy and faulty machine states, even under noisy conditions.
\end{abstract}


\section{Introduction}

\IEEEPARstart{R}{otating} machinery forms the core of many industrial systems, comprising components such as gears, rotors, and bearings that must operate continuously over long periods, for example, in wind turbines. These components are susceptible to faults that can cause unexpected failures, leading to unplanned downtime, safety risks, and substantial economic losses. As a result, condition monitoring and early fault diagnosis are essential to ensure the reliability, safety, and longevity of such machines \cite{zhang2010survey}.

A wide range of methods have been developed for condition monitoring and fault diagnosis of rotating machinery. Model-based approaches compare measured signals with outputs from physical system models \cite{lei2016model}, but their applicability is often limited by the complexity of real-world machines. In contrast, data-driven methods have proven more practical, leveraging historical vibration data to identify fault patterns \cite{yin2014review}. These approaches rely on signal processing and machine learning techniques to extract informative features from sensor measurements. While challenges such as data availability and sensor deployment remain, data-driven methods continue to offer a flexible and effective framework for fault diagnosis in industrial settings \cite{yin2014review}.

Discriminating between vibration signals from healthy and faulty machines is challenging, as raw time series data often contain redundant information, noise, and artifacts that obscure fault-related patterns. To address this, early approaches have extracted statistical features in the time and frequency domains \cite{jin2013motor,li2017independence}. While these features can capture certain aspects of signal behavior, their effectiveness is limited by the nonstationary and nonlinear nature of vibration signals, as well as by sensitivity to noise.

To overcome these limitations, time-frequency (T-F) analysis techniques have been widely adopted. Methods such as the short-time Fourier transform (STFT) \cite{he2017deep}, discrete wavelet transform (DWT) \cite{yan2014wavelets}, and empirical mode decomposition (EMD) \cite{lei2013review} enable localized signal
decomposition across both time and frequency. This allows for a more detailed characterization of signal components, especially in nonstationary settings, and improves the ability to isolate fault-relevant features. T-F representations thus serve as a powerful basis for identifying discriminative patterns in vibration signals for fault diagnosis.

Vibration signals from rotating machinery often exhibit complex fluctuations over a wide range of amplitudes and time scales, suggesting the presence of self-similar or fractal-like behavior \cite{liebovitch2003introduction}. This is particularly true when fault-induced vibrations are superimposed on
background noise and structural modes. To uncover these hidden patterns, detrended fluctuation analysis (DFA) \cite{peng1992long, xiong2017detrended} has been widely used as a tool to quantify long-range correlations and scale-invariant features in nonstationary signals \cite{kantelhardt2002multifractal,de2009applications}. DFA operates by isolating intrinsic fluctuations in the signal while eliminating low-frequency trends, thereby revealing the underlying correlation structure.

Extending this idea, multifractal detrended fluctuation analysis (MFDFA) characterizes not just one, but a spectrum of fluctuation behaviors across varying magnitudes and time scales. This enables the identification of multiple fractal patterns within the same signal, each associated with different fault-related dynamics. The resulting multifractal spectrum captures rich structural information about the vibration signal and has been shown to provide robust discriminative features for distinguishing between healthy and faulty machines
\cite{lin2013fault,du2019fault,liu2020novel,liu2015rolling}.

Despite its effectiveness, standard MFDFA methods \cite{de2009applications,lin2013fault,du2019fault} face two key limitations when applied to multichannel vibration data. First, they are typically designed for univariate time series and fail to account for cross-channel correlations that naturally arise in multichannel
sensor setups. Ignoring these dependencies can lead to incomplete or misleading analysis of fault-related dynamics. Second, raw vibration data often contains significant redundancy, noise, and irrelevant fluctuations, which obscure meaningful patterns and reduce the discriminative power of the extracted features \cite{mousavi2025enhanced}.

To address these challenges, we propose a novel framework that combines two key innovations: i) we introduce a fully multivariate generalization of MFDFA (FM-MFDFA) by defining a new fluctuation function based on a Mahalanobis distance-weighted $L_{pq}$ matrix norm. Unlike the existing multivariate MFDFA (MMFDFA) approach \cite{kantelhardt2002multifractal}, which treats each channel independently using Euclidean norms and ignores inter-channel correlations, our formulation explicitly captures cross-channel dependencies and variance biases. This allows for a more faithful representation of the joint dynamics across multiple sensor signals; ii) to focus on fault-relevant dynamics, we incorporate multivariate variational mode decomposition (MVMD) to decompose the signal into narrowband components and suppress noise
and irrelevant vibrations. The resulting framework provides a robust and interpretable basis for fault diagnosis using multichannel vibration data.

This article is organized as follows. Section II presents the necessary background theory and foundational concepts. Section III details the proposed methodology. Section IV reports the wind turbine fault diagnosis results with discussion on results in Sections V. Finally, conclusions are presented in Section VI with a look at future works.
\section{Background and Foundational Methods}
To lay the groundwork for our proposed fully multivariate MFDFA, we first review the standard MFDFA methodology and introduce the matrix norm concepts that underpin our formulation. This section also highlights key mathematical tools that will be generalized later in our framework.
\subsection{Detrended Fluctuation Analysis (DFA) and MFDFA}
Detrended fluctuation analysis (DFA) is a widely used technique for quantifying long-range correlations in nonstationary time series \cite{peng1992long,xiong2017detrended}. It works by subtracting local polynomial trends from overlapping signal segments and computing the residual fluctuation energy as a function of scale. Multifractal DFA (MFDFA) extends this approach by analyzing fluctuations at multiple magnitudes, parameterized by a moment order $q$, thereby revealing a spectrum of scaling behaviors. This multifractal spectrum provides a rich representation of underlying signal structure and has proven useful in applications such as fault diagnosis \cite{lin2013fault,du2019fault,liu2020novel,liu2015rolling}. To enable multivariate extension, we next reformulate MFDFA using matrix norms that allow structured representation of fluctuations across signal segments. This sets the foundation for our proposed fully multivariate MFDFA method.
\subsection{Matrix Norm and their Role in MFDFA}
Before detailing our proposed method, we formalize the use of matrix norms to compute fluctuation functions in MFDFA. Matrix norms allow us to generalize scalar norms to handle multidimensional signal segments and will later support the multivariate extension of the method.
\begin{definition}[$L_{pq}$ matrix norm]
	\label{def1}
	\textit{Let $\pmb{Z}\in\mathbb{R}^{N\times M}$ denote an $N\times M$ matrix containing observations $z_{i,j} \}_{i=1}^N \}_{j=1}^M$; $L_{pq}$ norm of $\pmb{Z}$ is defined as follows}
	\begin{equation}\label{Eq01}
		\small{||\pmb{Z}||_{pq}=||z_{i,j}\}_{i=1}^N\}_{j=1}^M||_{pq} = \left (\sum_{i=1}^{N} \Big(\sum_{j=1}^{M}\left(z_{i,j}\right)^p\Big)^\frac{q}{p}\right)^{\frac{1}{q}}.}
	\end{equation}
\textit{Proof that $||z_{n,m}\}_{n=1}^N\}_{m=1}^M||_{pq}$ fulfills all the conditions of a norm is available in} \cite{datta2010numerical}.
\end{definition}
This norm allows us to control how contributions from different rows and columns are aggregated, enabling flexible emphasis on specific fluctuation patterns.
\begin{remark}
	\label{remark1}
\textit{For $p=2$, $L_{pq}$ norm reduces to a special case $||\pmb{Z}||_{2q}$, as follows,}
\begin{equation}\label{Eq02} 
||\pmb{Z}||_{2q}=\left (\sum_{i=1}^{N} \Big(\sum_{j=1}^{M}\left(z_{i,j}\right)^2\Big)^\frac{q}{2}\right)^{\frac{1}{q}},
\end{equation}
\textit{that is an $L_q$ norm of the energies (i.e., $L_2$ norm) of the column vectors of $\pmb{Z}$. That means, for a given value of $q$, $||\pmb{Z}||_{2q}$ serves to emphasize vector observations with energies in a specific range, i.e., $||\pmb{Z}||_{2q}$ norm emphasizes column vectors with higher energy (or higher $L_2$ norm) for large $q$ while low energy column vectors are in focus when $q$ is smaller.}
\end{remark}
\subsection{Reformulating MFDFA using Matrix Norm Function}
To facilitate multivariate extension, we first express the steps of standard MFDFA using matrix norms. The key steps of the univariate MFDFA are as follows:
\begin{itemize}
    \item \textit{Cumulative Sum Profile:} Given a zero-mean time series $x_i \ \forall \ i=1,...,N$, the cumulative profile is computed as, $\small{y_i = \frac{1}{N}\sum_{j=1}^i (x_j-\frac{1}{N}\sum_{i=1}^N x_i)}$. This transformation amplifies long-term trends and fluctuations for more effective analysis.
    \item \textit{Segmentation:} The profile $y_i\}_{i=1}^N$ is divided into $L=N/s$ non-overlapping segments of length $s$, and mirrored to obtain $2L$ total segments, covering both ends of the series.
    \item \textit{Local Polynomial Detrending:} For each segment, a local polynomial fit $y_i$ (typically quadratic) is estimated via least-squares regression, and subtracted from the profile to extract residual fluctuations.
    \item \textit{Fluctuation Function:} The detrended segments are organized into a matrix $\mathbf{Z}\in\mathbb{R}^{2L\times s}$, where each row contains the fluctuations for one segment. The $q$-order fluctuation function is then computed using the $L_{2q}$ matrix norm as: 
    \begin{equation}\label{Eq03}
	\small{F_q(s) = \left (\frac{1}{2N_s}\sum_{v=0}^{2N_s-1} \Big(\frac{1}{s}\sum_{i=vs+1}^{(v+1)s}\left(y_i-\tilde{y}_i\right)^2\Big)^\frac{q}{2}\right)^{\frac{1}{q}}.}
\end{equation}
\end{itemize}
This formulation expresses the scale-dependent fluctuation magnitude at different orders $q$, allowing for multi-fractal analysis. Varying $q$ emphasizes different fluctuation magnitudes: large $q$ focuses on high-amplitude events, while negative $q$ highlights small-scale features.
\subsection{Multivariate MFDFA and its Limitations}
To enable analysis of multichannel time series, existing work \cite{kantelhardt2002multifractal} extends MFDFA using a three-dimensional matrix norm, defined as follows:
\begin{definition}[$L_{pqr}$ matrix norm]
	\label{def2}
	\textit{Let $\mathbf{Z}\in\mathbb{R}^{M\times N\times L}$ be a 3D matrix representing $N$ segments of multichannel signals, each of size $M\times L$. The $L_{pqr}$ norm is defined as:}
	\begin{equation}\label{Eq04}
		\small{||\pmb{Z}||_{pqr} = \left (\sum_{i=1}^{N} \Big(\sum_{j=1}^{M}\big(\sum_{k=1}^{L}(z_{i,j,k})^r\big)^\frac{p}{r}\Big)^\frac{q}{p}\right)^{\frac{1}{q}}}.
	\end{equation}
\end{definition}
This norm aggregates fluctuations across time ($L$), channels ($M$), and segments ($N$), is used to define a multivariate fluctuation function that generalizes the univariate MFDFA.
\begin{remark}
\label{remark2}
\textit{A common special case sets r = 2, yielding the Euclidean energy of the multichannel observations. This leads to what we denote as the $L_{pqr}$ norm, which can be interpreted as a matrix norm of the Euclidean distances (EDs) of the channel-wise fluctuations:
	\begin{equation}\label{Eq05}
	||\pmb{Z}||_{pq2} =\left (\sum_{i=1}^{N_s} \Big(\ \sum_{j=1}^{M}\Big(\sqrt{\pmb{z}_{i,j}^T \ \pmb{z}_{i,j}}\ \Big)^p\ \Big)^\frac{q}{p}\right)^{\frac{1}{q}} =||\pmb{Z}||_{pq^E},
\end{equation}	
where $\pmb{z}=\{z_1,\ldots,z_L\}$ and the scalar form of squared sums $(\sum_{l=1}^{L}(z_{l})^2)^\frac{1}{2}$ has been replaced by its vector form $\sqrt{\pmb{z}^T \ \pmb{z}}$. Essentially, $||\pmb{Z}||_{pq2}$ \eqref{Eq05} (or $L_{pq2}$ norm) is matrix norm of Euclidean distances (EDs) of the multichannel observations ${z}$ and hence we will refer to it as $L_{pq}$ norm of the EDs denoted by $L_{pq^E}=||\pmb{Z}||_{pq^E}=||\pmb{Z}||_{pq2}$.}
\end{remark}
This formulation enables a straightforward multivariate extension of the fluctuation function by aggregating fluctuation energy across time, channels, and segments. Specifically, the multivariate fluctuation function used in \cite{kantelhardt2002multifractal} is defined as:
\begin{equation}\label{Eq06}
	\small{F_{q^{M}}(s) = \left(\frac{1}{2N_s}\sum_{v=0}^{2N_s-1} \Big(\frac{1}{s}\sum_{i=vs+1}^{(v+1)s}\sum_{n=1}^{M}\left(y_{i_n}-\tilde{y}_{i_n}\right)^2\Big)^\frac{q}{2}\right)^\frac{1}{q},}
\end{equation}
where $y_{i_n}$ and $\tilde{y}_{i_n}$ denote the signal profile and local polynomial trend for the $n$-th channel in the $i$-th sample of segment $v$. This function measures the overall fluctuation energy across channels using a scalar Euclidean norm. However, this formulation suffers from a critical limitation: it treats each channel independently, simply summing their squared fluctuations without accounting for cross-channel relationships. In real-world sensor data, such as those used in vibration-based condition monitoring, sensor outputs are often strongly correlated due to shared mechanical structures, dynamic couplings, or external environmental influences. Neglecting these interdependencies can lead to an incomplete characterization of the joint behavior of the system and reduce the discriminative power of the extracted multifractal features. Therefore, while eq. \eqref{Eq06} and the MMFDFA approach in \cite{kantelhardt2002multifractal} extend MFDFA to multichannel settings, they fail to incorporate statistical dependencies between channels. In the next section, we address this limitation by introducing a fully multivariate generalization of MFDFA.
 \section{Proposed Method}
\label{sec:prop_method}
Building on the limitations identified in existing multivariate MFDFA formulations, we propose a fully multivariate generalization of MFDFA (FM-MFDFA) that incorporates cross-channel dependencies through a covariance-weighted matrix norm. Specifically, we replace the Euclidean norm in the fluctuation function with a Mahalanobis distance-based norm, which naturally accounts for both the correlation
structure and variance imbalance across sensor channels. This generalization allows the multiscale fluctuation function to accurately reflect the joint behavior of multichannel signals, a crucial aspect in vibration-based condition monitoring where signals from different sensors are often correlated. The resulting FM-MFDFA
framework retains the core structure of MFDFA while providing a more faithful representation of multivariate data. The key steps of our proposed method are as follows:

\subsection{Estimation of Inherent Multivariate Fluctuations}
The first step in FM-MFDFA is to isolate the intrinsic multichannel fluctuations in the signal by removing slow-varying trends that arise due to
nonstationarity. Let the multivariate time series be denoted as $x_i \in \mathbb{R}^M$, where $i= 1,\cdot,N$ and $M$ is the number of sensor channels.

We begin by constructing a cumulative profile that amplifies long-range correlations and suppresses high-frequency noise. This is computed as:
\begin{equation}\label{Eq07}
	\small{\pmb{y}_i = \frac{1}{N}\sum_{i=1}^N (\pmb{x}_i-\pmb{\overline{x}})}.
\end{equation}
The resulting profile ${y_i}_{i=1}^N$ is segmented into $L = \left \lfloor N/s \right \rfloor$ non-overlapping windows of equal length s, with mirrored segmentation to ensure full coverage.

To eliminate local trends, a least-squares polynomial fit of order $O$ is applied to each channel within every segment. Let $Y_v \in \mathbb{R}_{s\times M}$ denote the segment indexed by $v$, and let the fitted trend be denoted $\tilde{Y}_v$. The detrended signal is then:
\begin{equation}\label{Eq08}
    \hat{\mathbf{Y}}_v = \mathbf{Y} - \tilde{Y}_v
\end{equation}
Once all segments are detrended, the intrinsic multivariate fluctuations are reconstructed by concatenating all detrended segments:
\begin{equation}\label{Eq09}
    \hat{y}_1 = \Big [ \hat{Y}_1, \hat{Y}_2, ..., \hat{Y}_L \Big ]^T
\end{equation}
These fluctuations form the basis for computing the multiscale fluctuation function in the next step.
\subsection{Multivariate Fluctuation Function using Mahalanobis Norm}
Existing multivariate MFDFA approaches, such as \cite{kantelhardt2002multifractal}, rely on Euclidean norms to aggregate fluctuation energy across channels.
However, this assumption of statistical independence between sensor signals often fails in practice. In multichannel vibration data, cross-channel dependencies commonly arise due to physical coupling, shared loading conditions, and structural symmetries. Ignoring
these dependencies can result in incomplete or biased representations of signal dynamics. To address this, we propose a fully multivariate generalization of the fluctuation function using a Mahalanobis distance-based norm, which accounts for both variance imbalance and cross-channel correlations. The Mahalanobis norm generalizes the Euclidean norm by incorporating a covariance matrix that captures the statistical structure of the data.

Let $\mathbf{z}_{u,v} \in \mathbf{R}$ denote the detrended multivariate fluctuation vector at time index $v$ within segment $u$, and let
$\Sigma \in \mathbb{R}^{M\times M}$ be a positive definite covariance matrix estimated from the data. We define the following covariance-weighted norm:
\begin{definition}[Mahalanobis distance-based \texorpdfstring{$L_{pq}^\Sigma$}{Lpqsigma} norm]
	\label{definition3}
	\textit{Let $\mathbf{Z}\in\mathbb{R}^{U\times V\times M}$ be a 3D array of multivariate vectors $\mathbf{z}_{u,v}$. The covariance-weighted matrix norm is given by:}
\begin{equation}\label{Eq10}
	||\pmb{Z}||_{pq^\Sigma}=\left (\sum_{u=1}^{U} \Big(\ \sum_{v=1}^{V}\Big(\sqrt{\pmb{z}_{u,v}^T\ \Sigma^{-1} \ \pmb{z}_{u,v}}\ \Big)^p\ \Big)^\frac{q}{p}\right)^{\frac{1}{q}}
\end{equation}
\end{definition}
This norm extends the classical $L_{pq}$ norm formulation \eqref{Eq05} by applying Mahalanobis weighting to each fluctuation vector before aggregating across time and segment dimensions. Proofs that $||\pmb{Z}||_{pq^\Sigma}$ fulfills all the properties of a norm are given in Appendix A.

\begin{remark}[Interpretation under different covariance structures]
\textit{The behavior of this norm depends on the structure of $\Sigma$:
\begin{itemize}
\item When $\Sigma=I$, the formulation reduces to the standard Euclidean-based fluctuation function used in \cite{kantelhardt2002multifractal}.
\item When $\Sigma$ is diagonal with heterogeneous variances, it weights each channel based on its variability.
\item When $\Sigma$ contains off-diagonal terms, the norm accounts for cross-channel correlations, penalizing redundant or collinear fluctuations.
\end{itemize}}
\end{remark}
This flexibility allows the proposed method to generalize and improve upon earlier approaches by adapting to the underlying statistical structure of
the data. The proof of these cases is given in Appendix B.

Using this norm, we define the q-order fully multivariate fluctuation function as:
\begin{equation}\label{Eq11}
	F_q^{\Sigma}(s)=\left (\frac{1}{L}\sum_{l=1}^{L} \left(\ \frac{1}{s}\sum_{i=1}^{s}\hat{\pmb{y}}_{l,i}^T\ \Sigma^{-1} \ \hat{\pmb{y}}_{l,i}\ \right)^\frac{q}{2}\right)^{\frac{1}{q}}
\end{equation}
As in the standard in multifractal analysis, this expression becomes undefined when $q=0$. In that case, we compute the fluctuation function as the logarithmic mean, as follows
\begin{equation}\label{Eq12}
	F_0^{\Sigma}(s)=exp\left\{\frac{1}{2L}\sum_{l=1}^{L} ln \Big [\frac{1}{s}\sum_{i=1}^{s}\hat{\pmb{y}}_{l,i}^T\ \Sigma^{-1} \ \hat{\pmb{y}}_{l,i}\ \Big ] \right\}
\end{equation}

\begin{figure*}
	\centerline{\includegraphics[width=1.4\columnwidth]{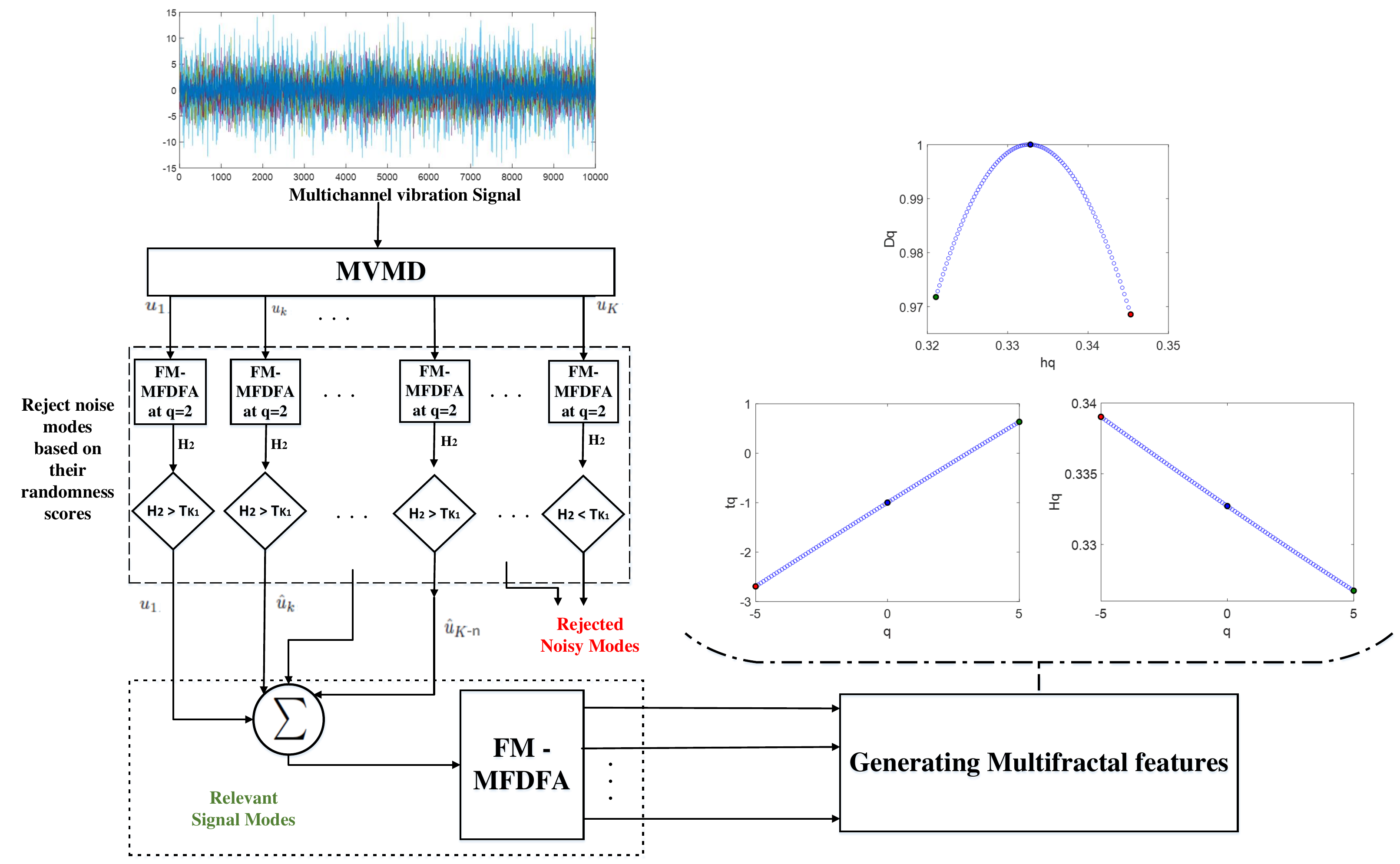}}
	\caption{Pipeline of the proposed fault diagnosis framework based on proposed GMDFA and GMF-MDFA}
	\label{fig01}
	\vspace{-5mm}
\end{figure*}

\subsection{Specifying fractal features}
To quantify the scale-dependent characteristics of the multivariate signal, we extract a set of standard multifractal features from the fluctuation function $F_q^{\Sigma}(s)$ defined in Eq. \eqref{Eq15}. These features capture the complexity, variability, and scaling behavior of the underlying signal across different magnitudes of fluctuation.

\paragraph{Generalized Hurst Exponent \texorpdfstring{$H_q$}{Hq}}
The first step is to estimate the generalized Hurst exponent $H_q$ by analyzing the power-law relationship between the fluctuation function and the scale parameter $s$. Specifically, for each moment order $q$, we fit the following relation:
\begin{equation}\label{Eq13}
	F_q^\Sigma(s) = s^{H(q)}.
\end{equation}
The exponent $H_q$ characterizes the correlation structure of fluctuations at different magnitudes:
\begin{itemize}
    \item $H_q > 0.5$ indicates long-range correlations,
    \item $H_q < 0.5$ suggests anti-persistent behavior, and
    \item Variation of $H_q$ with respect to $q$ reflects multi-fractal scaling.
\end{itemize}
\paragraph{Generalized Hurst Exponent \texorpdfstring{$H_q$}{Hq}}
From the estimated $H_q$, we derive the mass exponent $\tau_q$, which describes how the $q$-th order statistical moment of the signal scales with resolution:
\begin{equation}\label{Eq14}
	\tau_q = q H_q - 1,
\end{equation}
This function is central to characterizing multifractality. For monofractal signals, $\tau_q$ is linear in $q$, whereas nonlinear $\tau_q$ curves indicate
multifractal behavior.
\paragraph{Singularity Spectrum \texorpdfstring{$f(\alpha_q)$}{falphaq}}
Finally, we compute the singularity spectrum which provides a geometric interpretation of the signal's local regularity. It is derived
from $\tau_q$ using a Legendre transform:
\begin{equation}\label{Eq15}
	f(\alpha_q) = q\alpha_q - \tau_q = q(\alpha_q - H_q)+1.
\end{equation}
Here:
\begin{itemize}
    \item $\alpha_q = H_q - q H_q^{'}$ is the Holder exponent, describing the strength of singularities.
    \item $f(\alpha_q)$ is the dimension of the subset of the signal that exhibits singularity $\alpha_q$.
\end{itemize}
A wider spectrum implies greater signal complexity and richer dynamics, which often correspond to the presence of faults or abnormal conditions in vibration signals.

\subsection{Fault Diagnosis Application to Wind Turbine Vibration Data}
Having established the FM-MFDFA framework for analyzing multivariate multifractal fluctuations, we now shift our focus from methodological formulation to practical application. The goal of this section is to demonstrate the effectiveness and interpretability of the proposed method for fault diagnosis in rotating machinery using real vibration sensor data.

We apply FM-MFDFA to multichannel vibration recordings obtained under both healthy and faulty operating conditions. This allows us to evaluate how well the extracted multifractal features capture fault-relevant signal dynamics and distinguish between machine states. In particular, we assess whether the dependency-aware fluctuation analysis enabled by FM-MFDFA offers improved sensitivity
compared to conventional univariate or Euclidean-based approaches.

The overall workflow of the proposed fault diagnosis pipeline is illustrated in Figure 1. It consists of three key stages: (i) preprocessing and segmentation of multichannel vibration signals, (ii) extraction of multifractal features using FM-MFDFA, and (iii) analysis and comparison of feature distributions across health states.

\subsection{MVMD based estimation of discriminant features}
To isolate fault-relevant information and suppress noisy or redundant components, we incorporate multivariate variational mode decomposition (MVMD) \cite{ur2019multivariate} prior to fractal analysis. MVMD jointly decomposes a multichannel vibration signal $\mathbf{x}_i\in \mathbb{R}^M$ into $K$
narrowband intrinsic mode functions (IMFs) $\mathbf{u}_i\in \mathbb{R}^M$, each associated with a central frequency $\omega_k$. The decomposition is obtained by
minimizing the joint bandwidth of the modulated components, as follows
\begin{equation}\label{Eq16}
	\small{\mathop{\text{minimize}}_{u_{k,n},w_k}\left \{ \sum_{k=1}^K\sum_{n=1}^m\left \| \partial_t \left[u_{k,i_n}^{+}\ e^{-jw_k i}\right] \right \|_2^2\right \},}
\end{equation}
subject to the constraint that
\begin{equation} \label{Eq17}
	\small{\pmb{x}_{i}=\sum_{k=1}^K \pmb{u}_{k,i}}.
\end{equation}
This ensures that the IMFs are band-limited and jointly estimated across channels, preserving cross-sensor dependencies and reducing mode mixing.
\begin{figure*}[t]
	\centering
	\begin{subfigure}{.23\textwidth}
		\includegraphics[width=\linewidth]{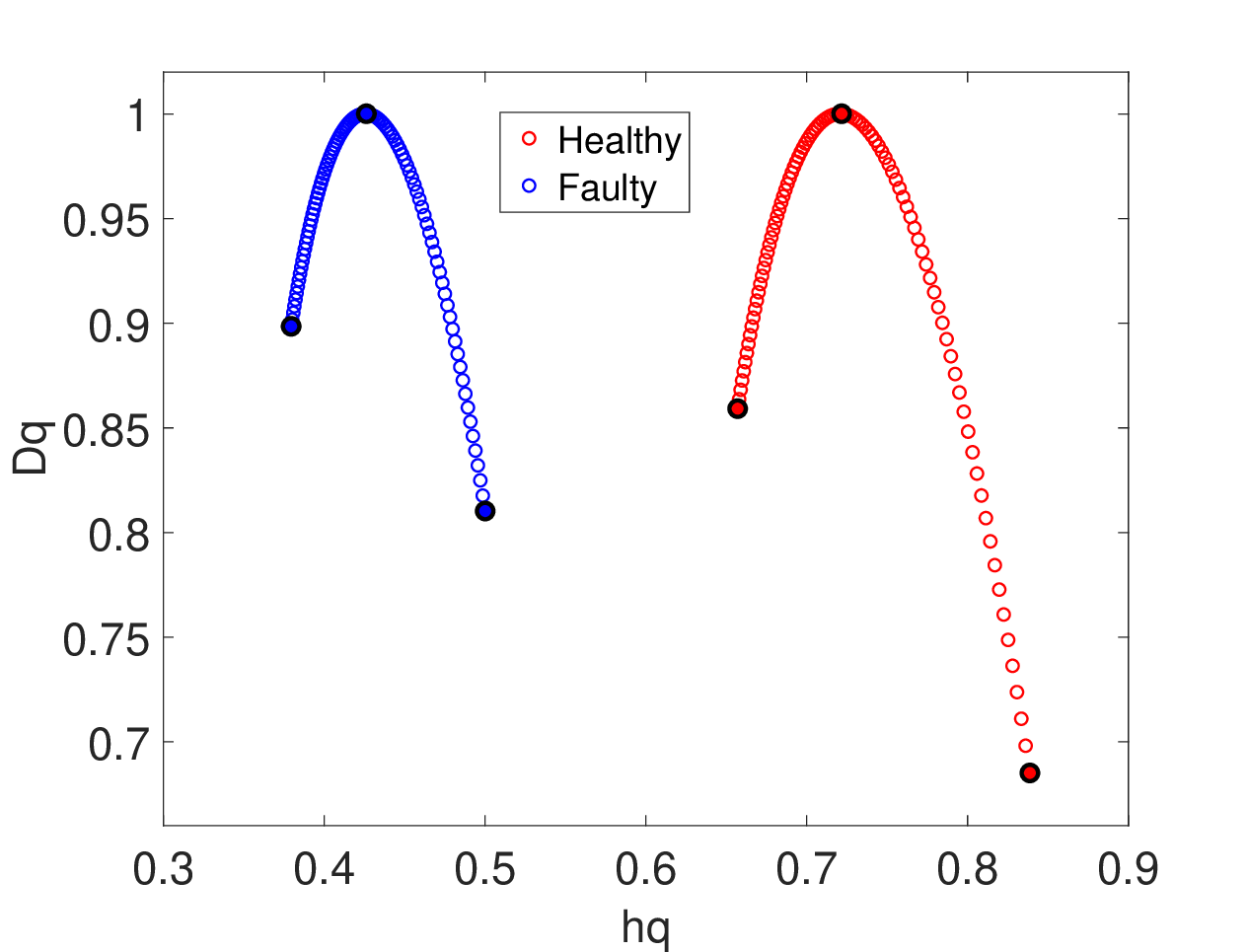}
		\caption{IMF1}
		\label{fig2a}
	\end{subfigure}
	\begin{subfigure}{.23\textwidth}
		\includegraphics[width=\linewidth]{Fig_imf1.eps}
		\caption{IMF2}
		\label{fig2b}
	\end{subfigure}
	\begin{subfigure}{.23\textwidth}
		\includegraphics[width=\linewidth]{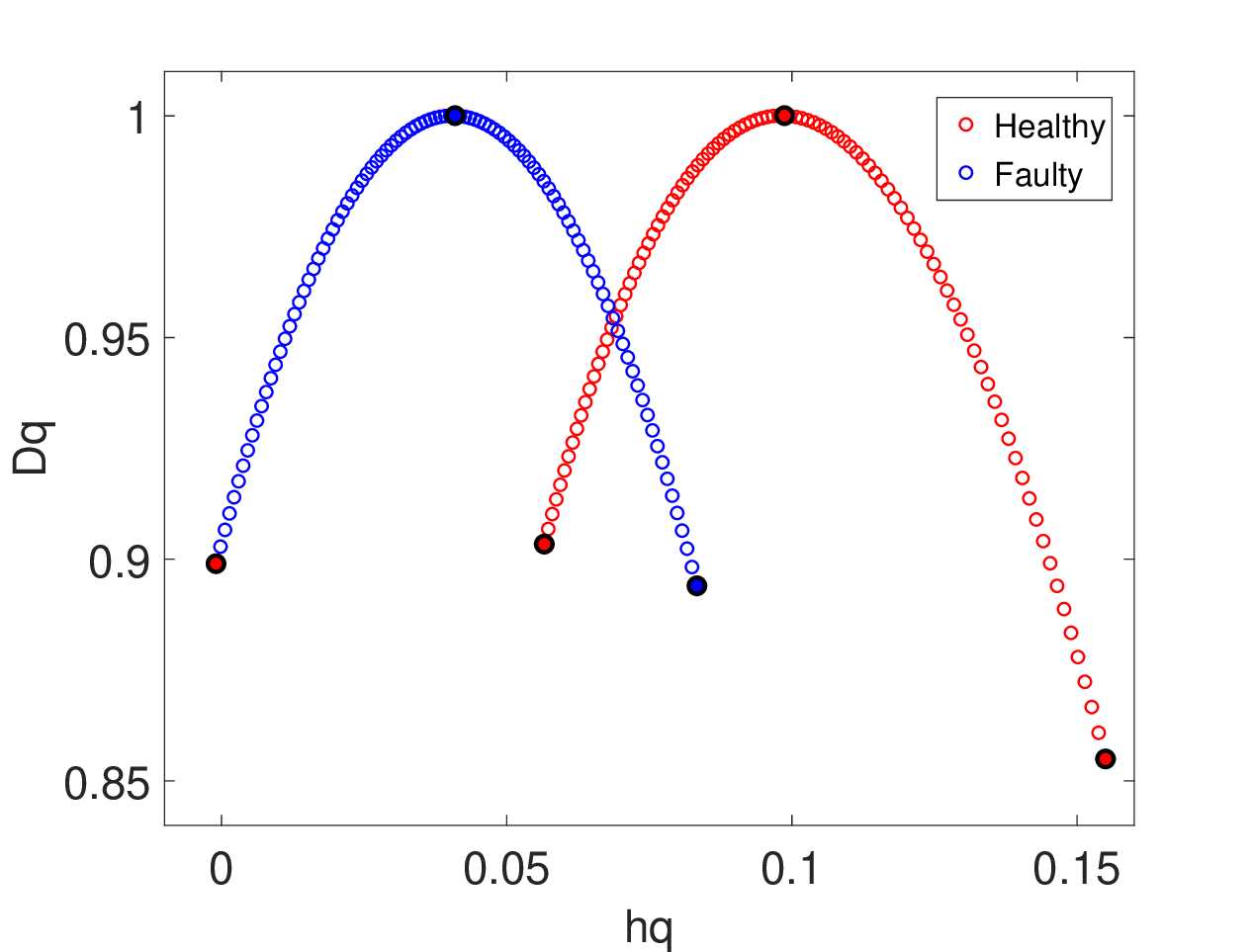}
		\caption{IMF3}
		\label{fig2c}
	\end{subfigure}
	\begin{subfigure}{.23\textwidth}
		\includegraphics[width=\linewidth]{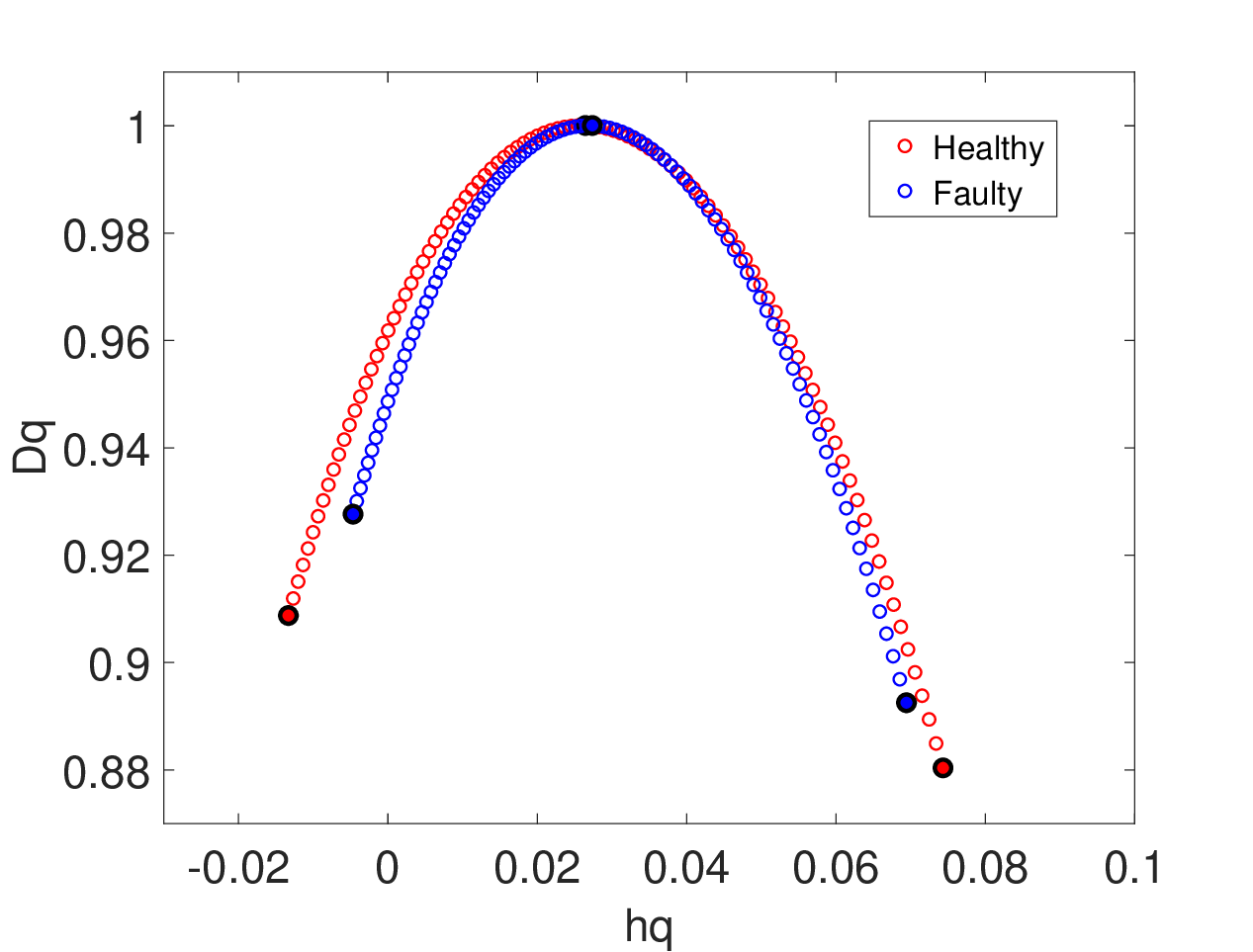}
		\caption{IMF4}
		\label{fig2d}
	\end{subfigure}
	\begin{subfigure}{.23\textwidth}
		\includegraphics[width=\linewidth]{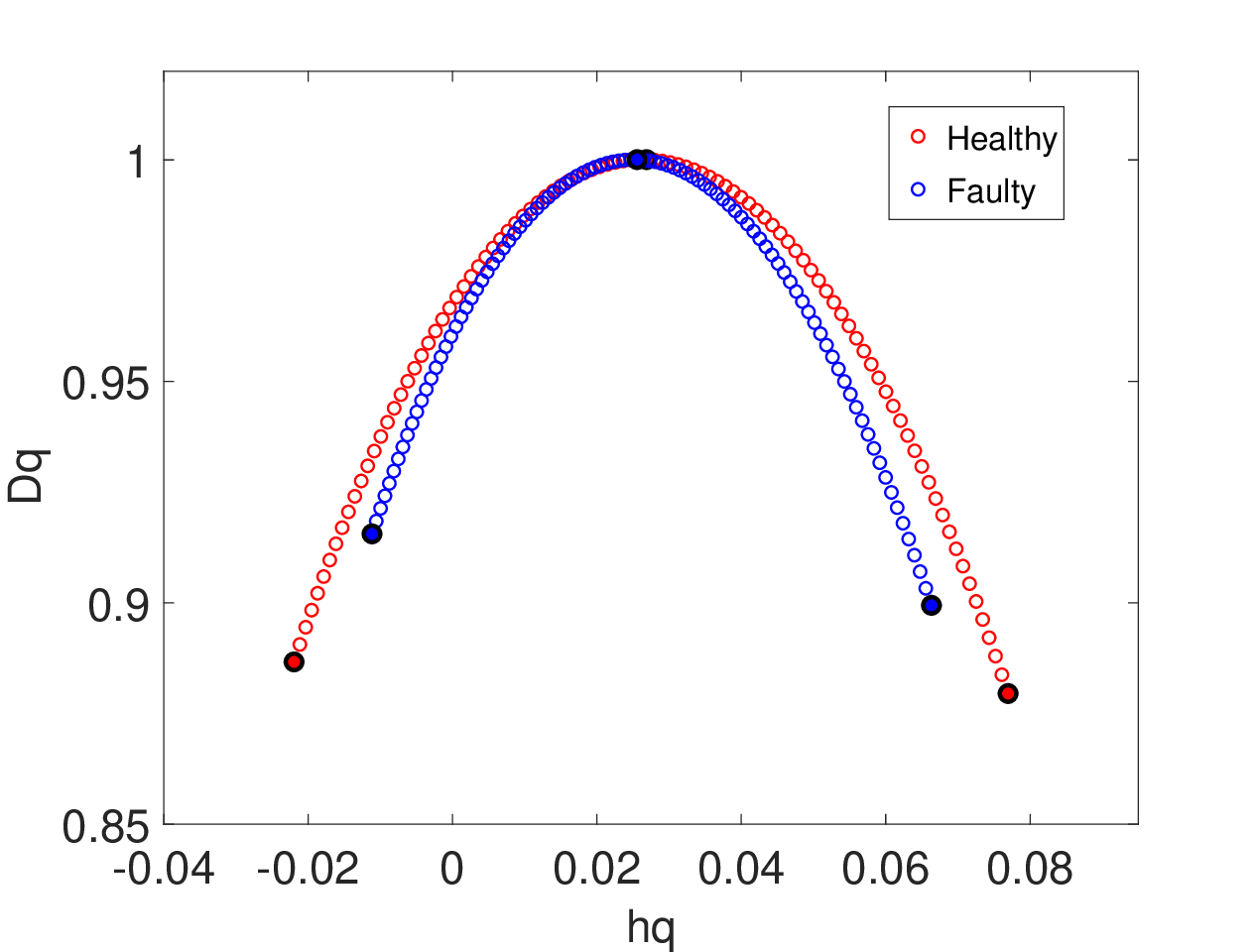}
		\caption{IMF5}
		\label{fig2e}
	\end{subfigure}
	\begin{subfigure}{.23\textwidth}
		\includegraphics[width=\linewidth]{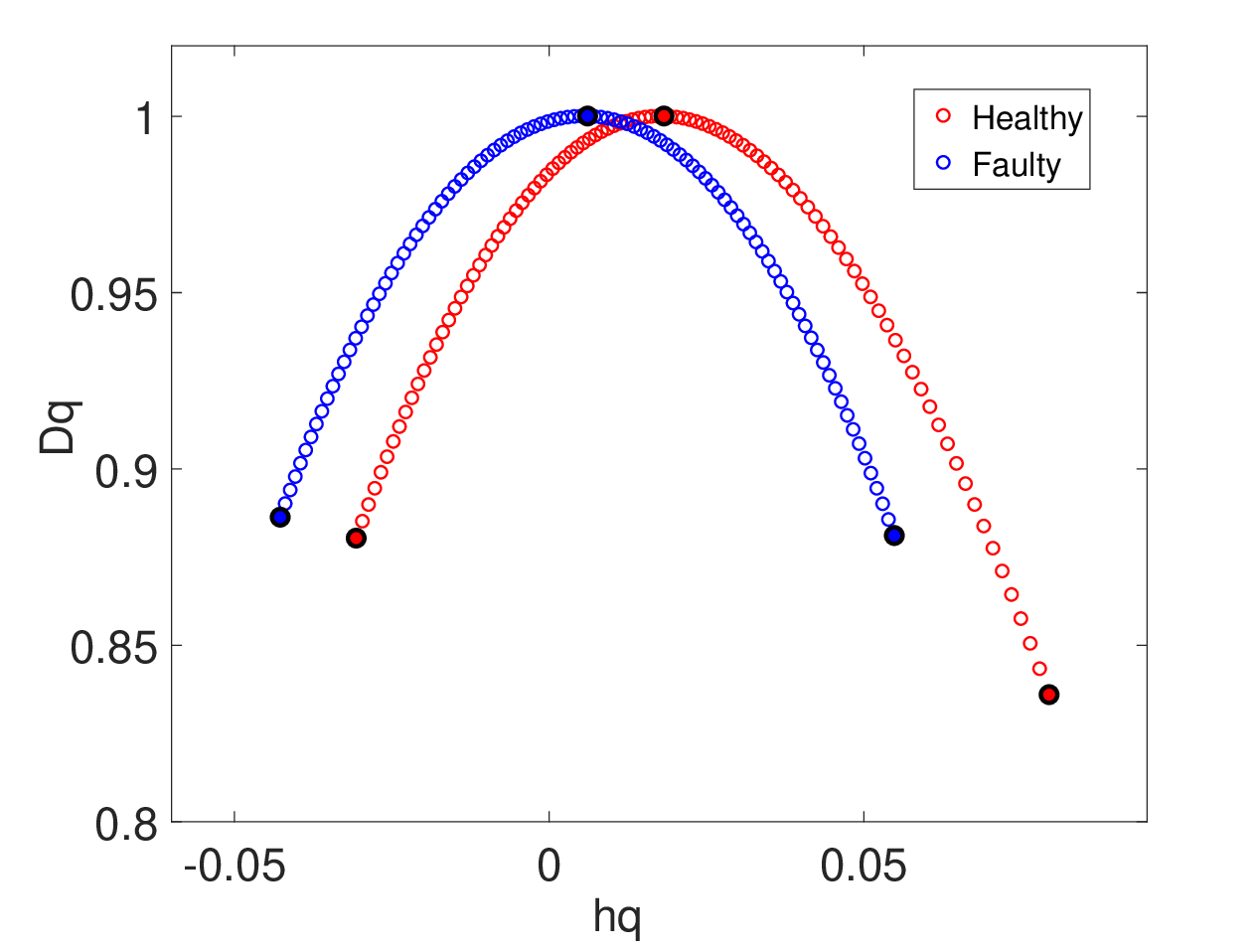}
		\caption{IMF6}
		\label{fig2f}
	\end{subfigure}
	\begin{subfigure}{.23\textwidth}
		\includegraphics[width=\linewidth]{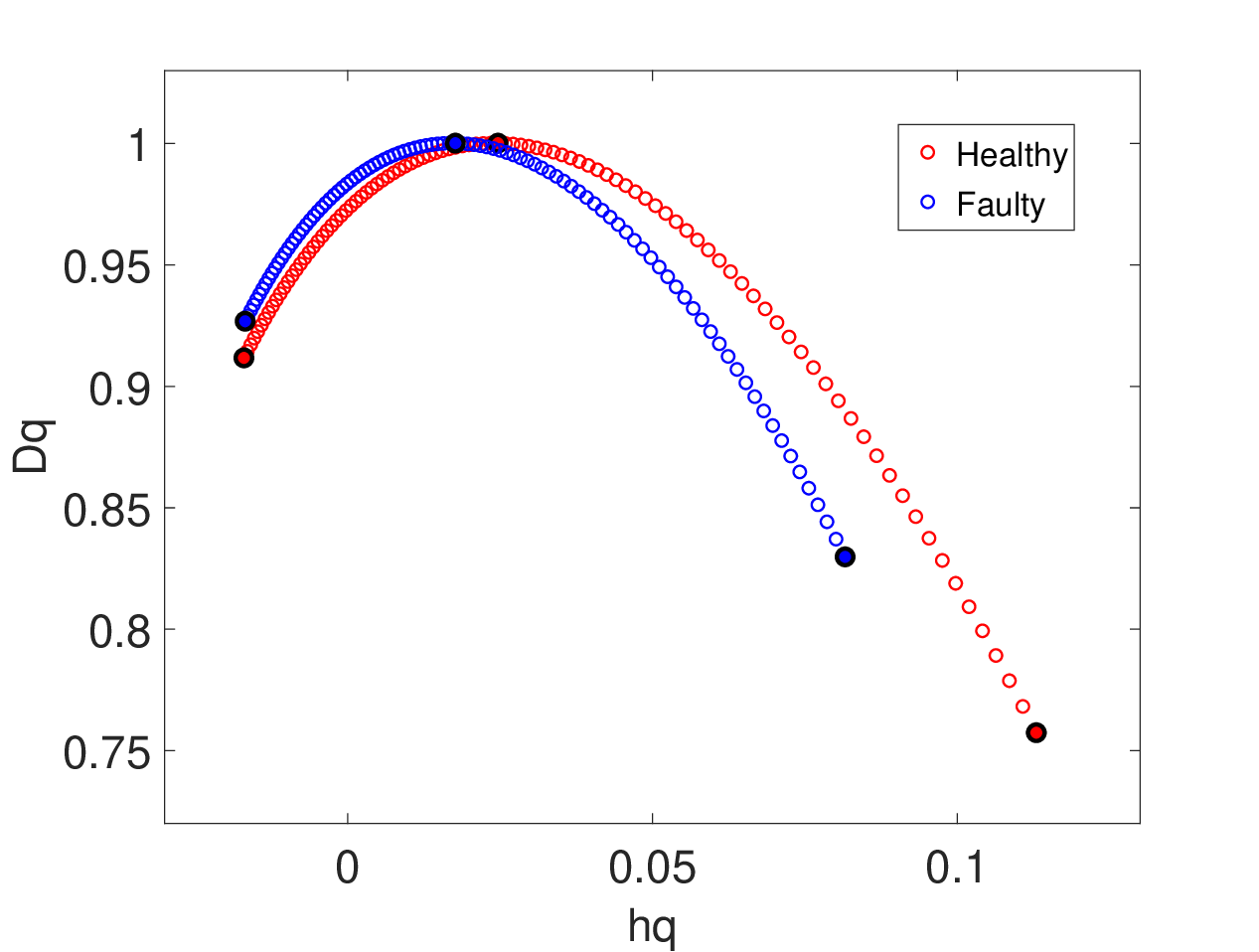}
		\caption{IMF7}
		\label{fig2g}
	\end{subfigure}
	\begin{subfigure}{.23\textwidth}
		\includegraphics[width=\linewidth]{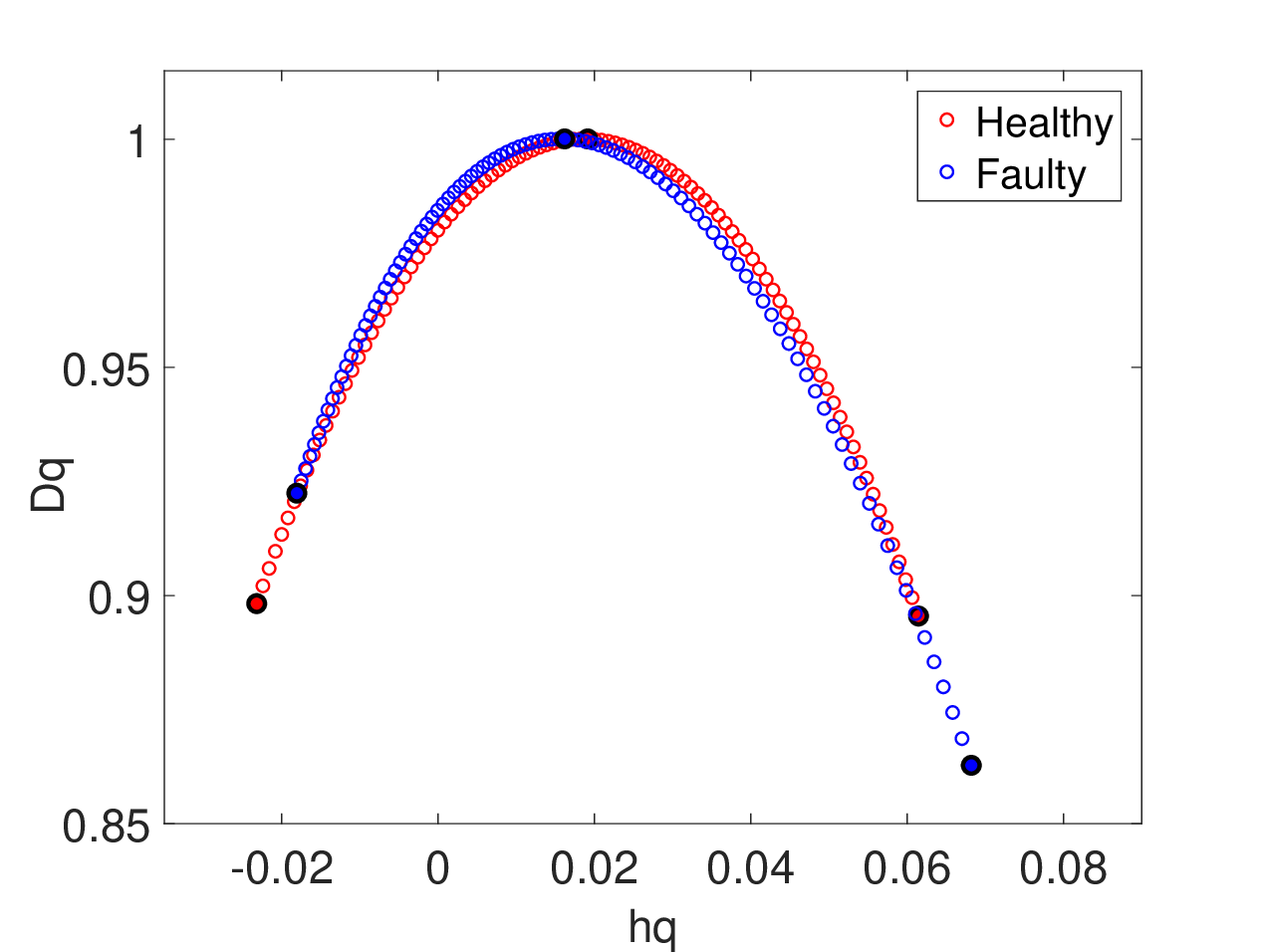}
		\caption{IMF8}
		\label{fig2h}
	\end{subfigure}
	\caption{Comparison of fractal spectrum of individual MVMD modes of multichannel vibration data from a healthy and a faulty machines, labeled \texorpdfstring{$H_1$}{H1} and \texorpdfstring{$D_1$}{D1} in \cite{sheng2014wind}.}
	\label{fig2}
	\vspace{-5mm}
\end{figure*}

To focus analysis on fault-relevant components, we separate the IMFs into two groups, as follows
\begin{itemize}
    \item Low-order modes $\{\pmb{u}_{k,i}\}_{k=1}^{K_1}$, assumed to contain deterministic signal structure and fault-related information.
    \item High-order modes $\{\pmb{u}_{k,i}\}_{k=K_1+1}^{K}$, assumed to represent noise and irrelevant fluctuations.
\end{itemize}
The signal is thus partitioned as
\begin{equation}\label{Eq18}
	\small{\pmb{x}_{i} =\sum_{k=1}^{K_1} \pmb{u}_{k,i} + \sum_{k=K_1+1}^K \pmb{u}_{k,i}}, \ \forall\ i=1,\ldots,N,
\end{equation}
To identify $K_1$, we leverage the Hurst exponent $H_2$ (computed via FM-MFDFA with $q = 2$), which measures long-range temporal correlations. At $q=2$, proposed FM-MFDFA reduced to the generic multivariate DFA (GMDFA) presented in our previous works in \cite{naveed2025FMDFA,naveed2021GMDFAconf}, where we have already shown its efficacy in estimating and reconstructing signal without noise. Moreover, our work in \cite{naveed2021statistical,naveed2024VMDTDIP,naveed2024variational} details different ways to select IMFs corresponding to signal of interest in noisy conditions. IMFs with low $H_2$ indicate randomness, while those with higher $H_2$ retain structural dependencies. We compute the difference between successive Hurst exponents and select the cutoff point where the largest change occurs, as follows
\begin{equation}\label{Eq19}
	\small{K_1 = \mathop{\text{argmax}}_{k}\left( \left \{|H_2^k-H_2^{k+1}| \right \}_{k=1}^{K-1} \right )}
\end{equation}
The reconstructed fault-relevant signal is then given by
\begin{equation}\label{Eq20}
	\hat{\pmb{x}}_{i} =\sum_{k=1}^{K_1} \pmb{u}_{k,i}, \ \forall\ i=1,\ldots,N.
\end{equation}
This signal is subsequently used as input to FM-MFDFA to extract multifractal features, as described in the next section.

\subsection{FM-MFDFA based fault estimation}
We apply the FM-MFDFA procedure (as defined in \eqref{Eq11}) to the reconstructed signal $\hat{\pmb{x}}_{i}$ obtained by summing the most informative MVMD
modes \eqref{Eq18}. This allows us to extract scale-dependent multifractal features that reflect the underlying dynamics of machine vibrations.
Following the FM-MFDFA pipeline, we estimate the generalized Hurst exponents $H_q$ from the power-law scaling of the fluctuation function with
respect to segment size $s$ \eqref{Eq14}, and compute the corresponding mass exponent $\tau_q$, singularity strength $\alpha_q$, and multifractal spectrum $f(\alpha_q)$ using Equations \eqref{Eq13}-\eqref{Eq15}. From these, we construct a final feature vector $\Phi_{FM}$, including descriptors such as:
\begin{itemize}
    \item Hurst spectrum width $\Delta H = H_{max} - H_{min}$,
    \item Singularity spectrum width and skew,
    \item Peak singularity strength $\alpha_{peak}$,
    \item and other moments or shape descriptors.
\end{itemize}
These features form the basis for visual analysis, clustering, or supervised classification of machine health states in the subsequent
experiments.

Since the vibration data from healthy and faulty gearboxes contain vibrations of different natures, their fractal spectrum should be significantly far from each other. Study of these differences can lead to development of a way to classify between the healthy and faulty machines in the fractal space. Hence, a statistical distance function may be used to estimate the differences in the fractal features of the healthy and faulty machine data. Let $\mathcal{D}_{statistics}$ denotes a statistical distance metric from the origin, that is applied on the fractal features obtained in the previous step to get a measure of 
\begin{equation}
	D_{MF} = \mathcal{D}_{statistics} \{\tau_q,H_q,f_{\alpha_q}\},
\end{equation}
where $D_{MF}$ denotes a multichannel statistical distance, e.g., Euclidean distance, Mahalanobis distance etc., of the multi-fractal (MF) features. 

Similar nature of the fluctuation and magnitude patterns in the vibration data from healthy machines dictates that their fractal spectrum should be similar too. Hence, a measure of the maximum variation in the fractal spectrum of healthy machines along with their distance from the fractal features of the faulty machines can lead to estimation of a threshold $T_{h-d}$, as follows
\begin{equation}
	\{\tau_q^h,H_q^{h},f_{\alpha_q}^h\} \xleftrightarrow{\text{Threshold $T_{h-d}$}} \{\tau_q^d,H_q^{d},f_{\alpha_q}^d\},
\end{equation}
where $\{\tau_q^h,H_q^h,f_{\alpha_q}^h\}$ and $\{\tau_q^d,H_q^{d},f_{\alpha_q}^d\}$ respectively denotes fractal features from healthy and faulty machines vibration data $\pmb{x}_i^h\}_{i=1}^N$ and $\pmb{x}_i^d\}_{i=1}^N$. Since threshold value $T_{h-d}$ is a measure of the deviation from normal fractal behavior and minimum distance from the faulty behaviour, an aberration from the normal behaviour  i.e., $D>T_{h-d}$, means detection faulty machine, while the case where $D<T_{h-d}$ suggests detection normal activity, i.e., healthy machine. 

In this work, we use visual illustrations to show the difference between the two vibration datasets in the fractal space. This way, we not only show the efficacy of the proposed approach but also keep the manuscript compact. Results below demonstrate that the fractal analysis using the proposed FM-MFDFA method can fully  between the datasets from healthy and faulty machines.
\begin{figure*}[t]	
	\centering
	\begin{subfigure}{.265\textwidth}
		\includegraphics[width=\linewidth]{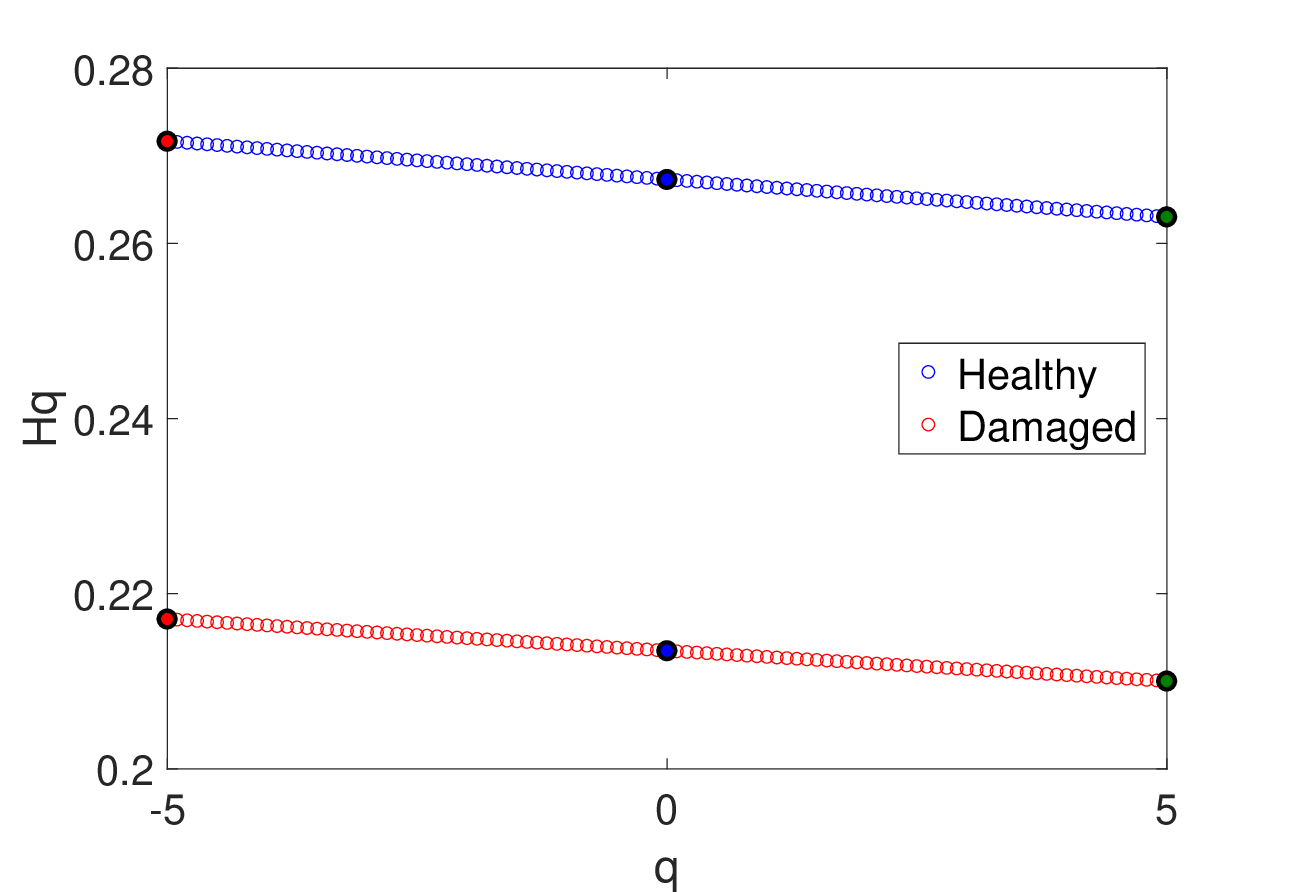}
				\caption{}
		\label{fig3a}
	\end{subfigure}
	\hspace{-4mm}
	\begin{subfigure}{.265\textwidth}
		\includegraphics[width=\linewidth]{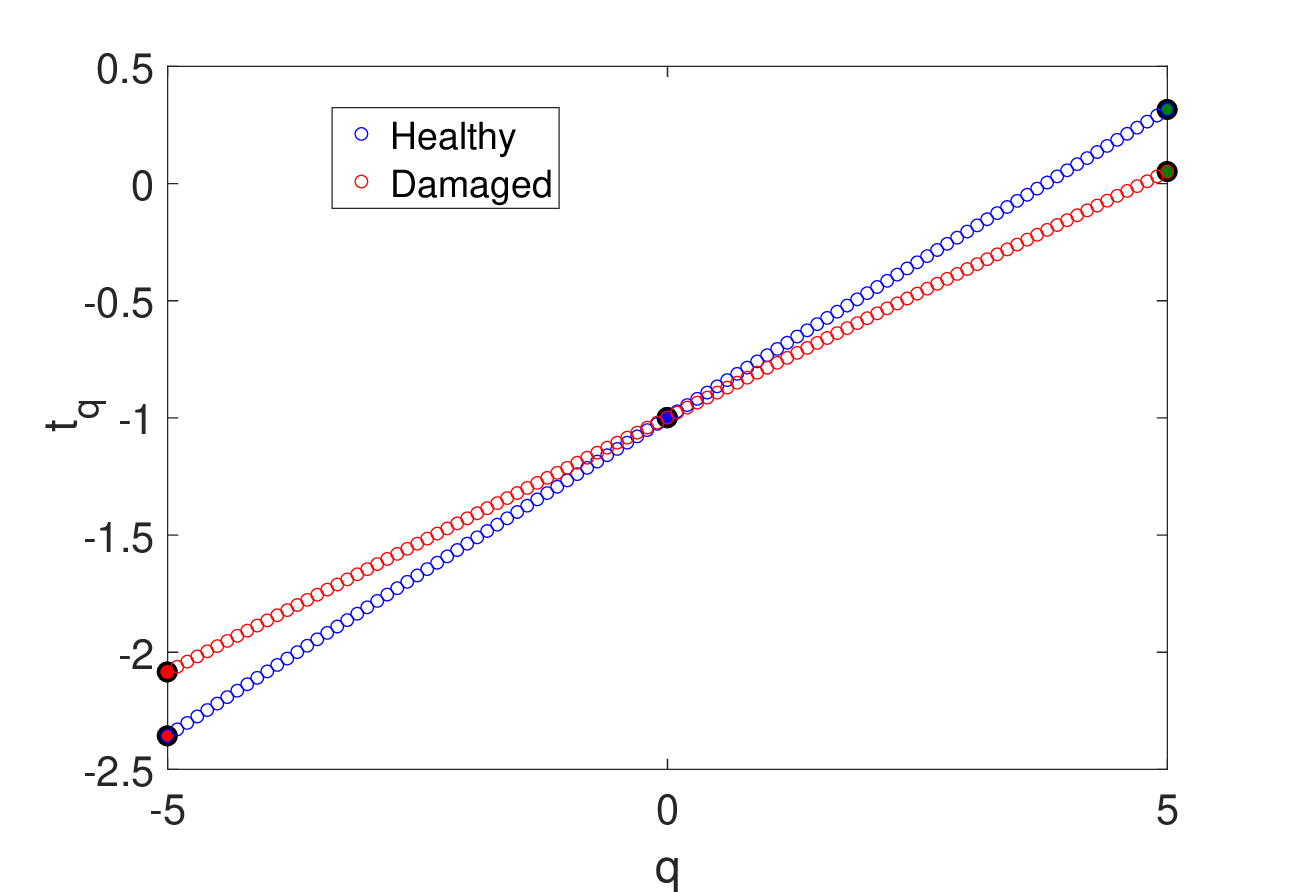}
				\caption{}
		\label{fig3b}
	\end{subfigure}
	\hspace{-4mm}
	\begin{subfigure}{.265\textwidth}
		\includegraphics[width=\linewidth]{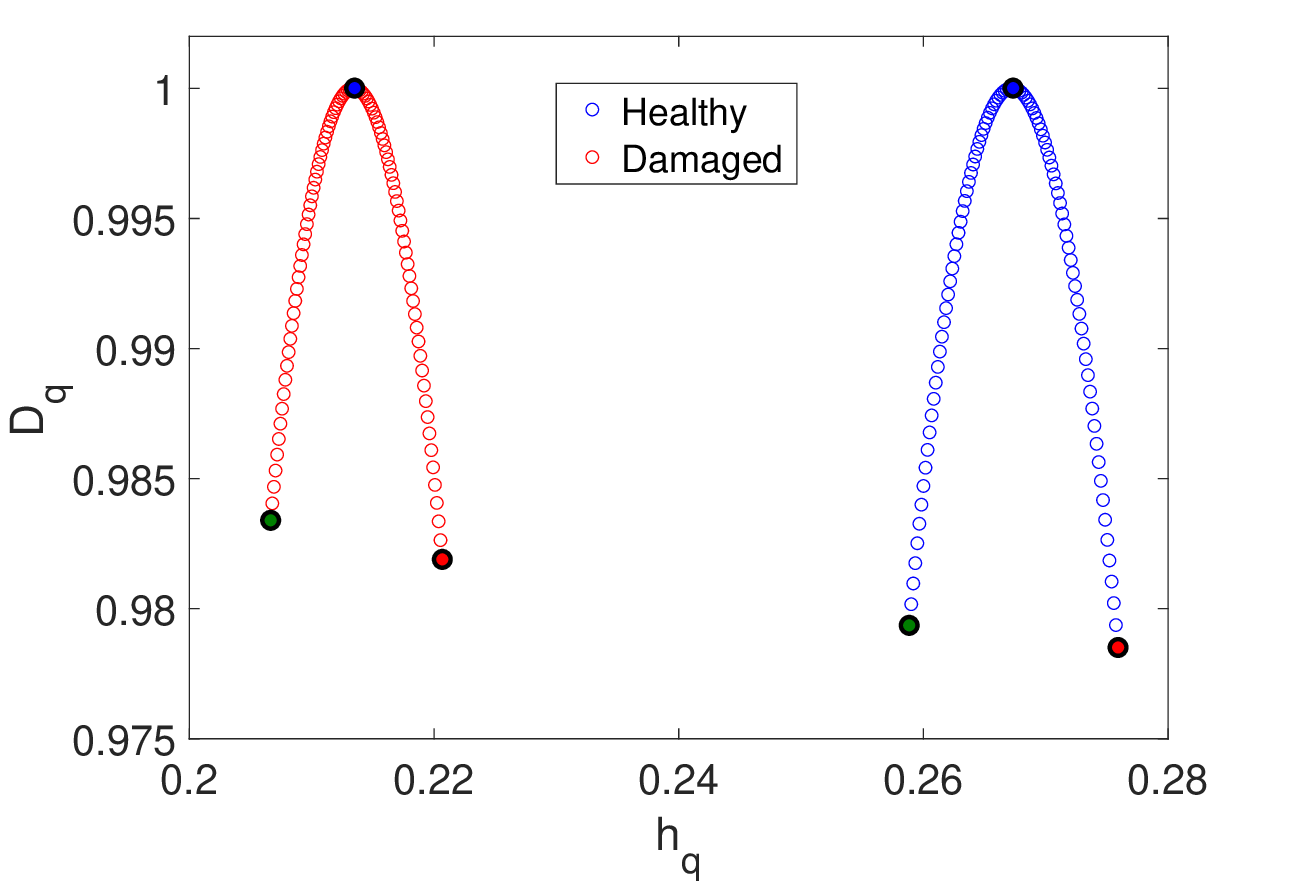}
				\caption{}
		\label{fig3c}
	\end{subfigure}
	\caption{A comparison of the reconstructed fractal features from the first three MVMD modes of healthy and damaged Wind Turbines (labeled \texorpdfstring{$H_1$}{H1} and \texorpdfstring{$D_1$}{D1}), plotted in Fig. \ref{fig2} Turbines.}
	\label{fig3}
\end{figure*}
\begin{figure*}[t]	
	\centering
	\begin{subfigure}{.265\textwidth}
		\includegraphics[width=\linewidth]{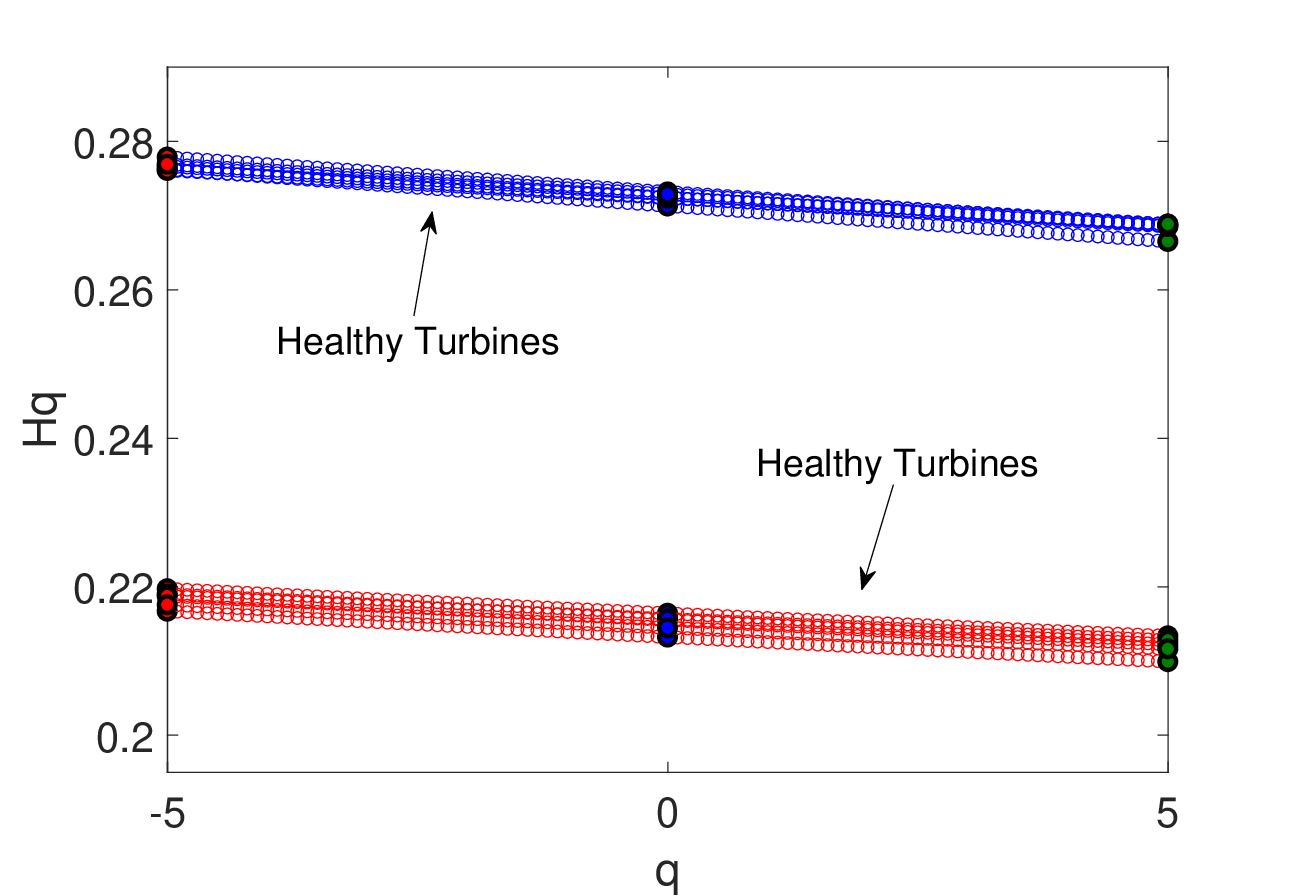}
		\caption{}
		\label{fig4a}
	\end{subfigure}
	\hspace{-4mm}
	\begin{subfigure}{.265\textwidth}
		\includegraphics[width=\linewidth]{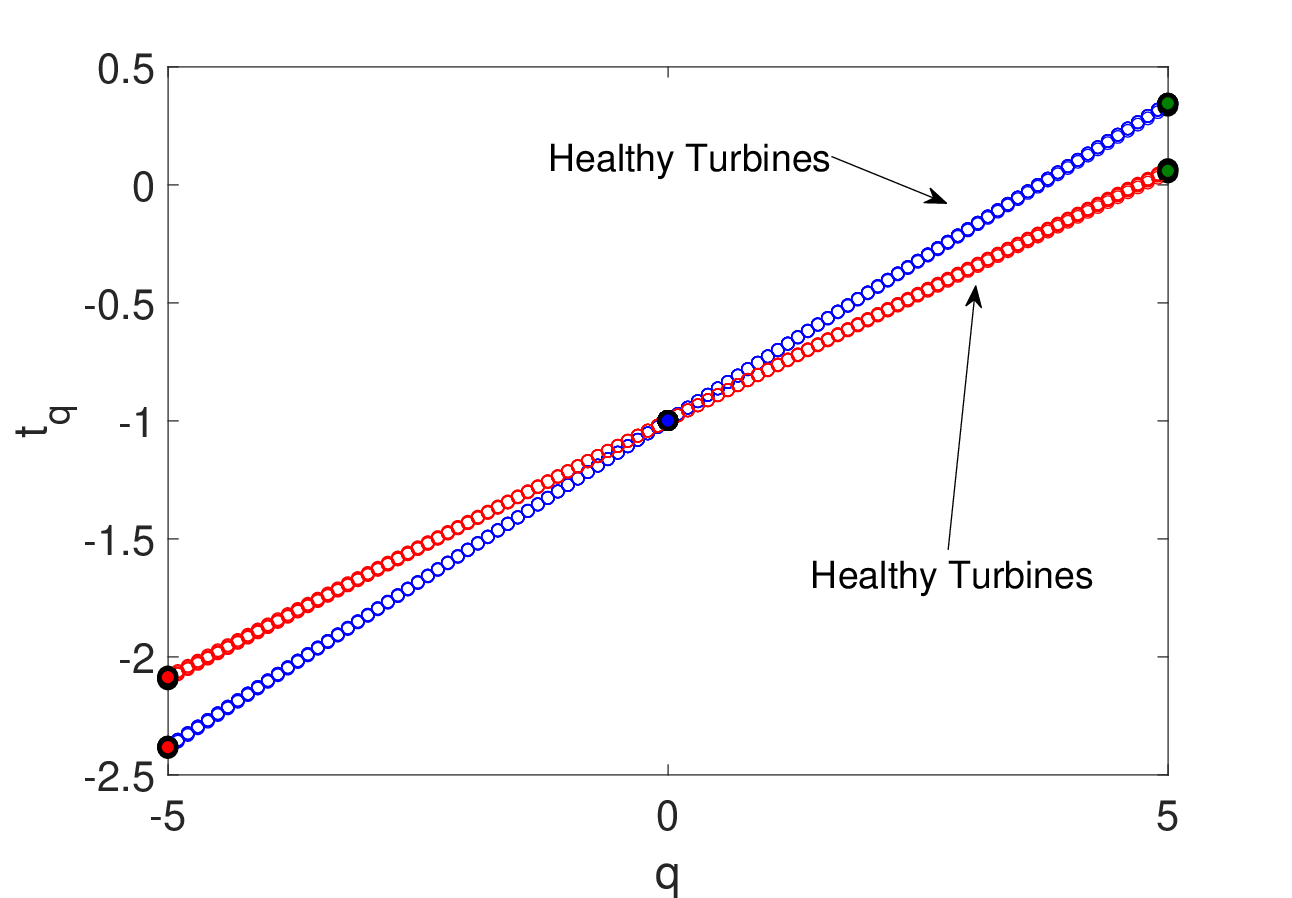}
		\caption{}
		\label{fig4b}
	\end{subfigure}
	\hspace{-4mm}
	\begin{subfigure}{.265\textwidth}
		\includegraphics[width=\linewidth]{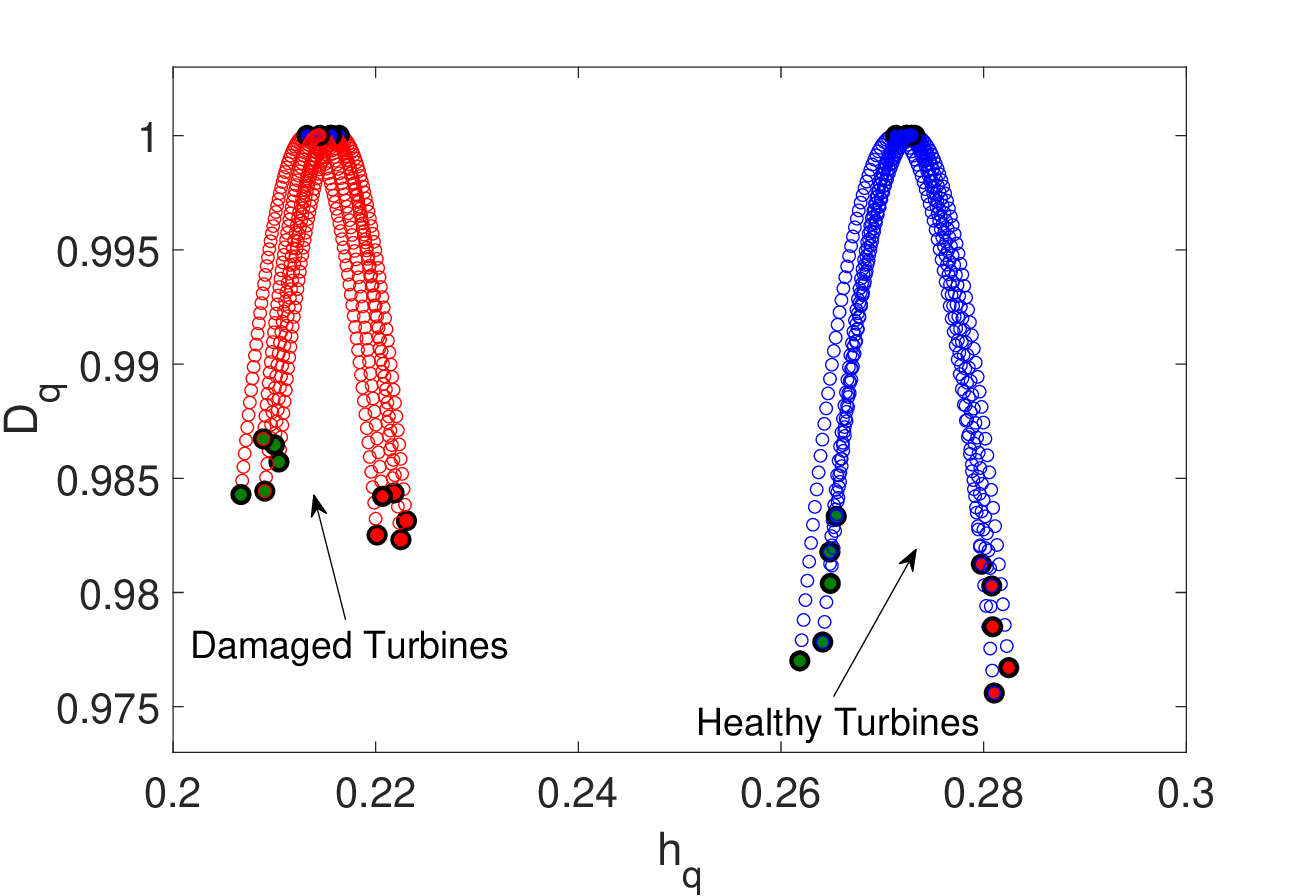}
		\caption{}
		\label{fig4c}
	\end{subfigure}
	\caption{Plot of scaling exponents \texorpdfstring{$\alpha_k$}{alphak}, computed using proposed FM-MFDFA at \texorpdfstring{$q=2$}{q=2}, for MVMD modes of noisy Wind signal at \texorpdfstring{$10$}{10}dB.}
	\label{fig4}
\end{figure*}
\section{Results}

This section demonstrates that the fractal spectra of healthy and faulty machines, obtained using the proposed FM-MFDFA method, are clearly distinguishable. This distinction enables reliable fault identification by observing whether the vibration behavior of a gearbox deviates from that of a healthy machine within the fractal feature space. To validate the performance of the proposed fault-diagnosis approach, we evaluate it on vibration data from wind turbines. Specifically, we use the NREL wind turbine vibration dataset \cite{sheng2014wind}, collected from \textit{healthy} and \textit{damaged} gearboxes of the same design during the Wind Turbine Gearbox Reliability Collaborative (GRC) dynamometer tests. Accelerometers and high-speed shaft RPM sensors were employed to record vibratory responses during operation. The accelerometers were mounted on the exterior of the gearbox to capture surface vibratory acceleration. Each channel was sampled at $40$ kHz for $10$ minutes using a high-speed data acquisition system (DAQ). Eight sensors, labeled AN3--AN10, were used to simultaneously record vibrations from multiple gearbox components. Further details regarding the sensor configuration can be found in \cite{sheng2014wind}.

We begin by comparing the fractal spectra $f(\alpha_q)$ of the intrinsic mode functions (IMFs) obtained by applying MVMD to the vibration data from a healthy and a faulty gearbox, labeled $H_1$ and $D_1$, respectively, as shown in Fig.~\ref{fig2}. These results demonstrate not only the effectiveness of MVMD in isolating the most discriminative components, but also the ability of the proposed FM-MFDFA method to clearly separate IMFs containing fault-related dynamics from those dominated by noise. Notably, only a few of the initial IMFs exhibit significant differences between healthy and faulty conditions. Figures~\ref{fig2a}--\ref{fig2c} show that the fractal spectra of the first three IMFs are highly distinguishable, whereas the fractal behaviour of the healthy and faulty machines in the later IMFs (Figs.~\ref{fig2d}--\ref{fig2h}) are nearly identical. This indicates that the first three IMFs capture the majority of fault-relevant features, while subsequent IMFs are primarily random noise and contribute little to machine-health characterization.

The reconstruction of the fractal spectra using only the first three IMFs is shown in Fig.~\ref{fig3a}. Figures~\ref{fig3b} and \ref{fig3c} further compare other multifractal features, specifically the scaling exponents $H_q$ and $\tau_q$ as functions of the order $q$. As discussed previously, $H_q$ characterizes the scaling behavior of fluctuations at different magnitudes, while $\tau_q$ and $f(\alpha_q)$ describe the multifractal structure. These figures confirm substantial differences between the multifractal properties of the healthy and faulty datasets. The differences in slopes of the $\tau_q = qH_q - 1$ curves (Fig.~\ref{fig3b}) reflect distinct scaling mechanisms, while the fractal spectra in Fig.~\ref{fig3c} highlight the divergence in local regularity over time. The fractal spectra of the two conditions are well separated: the dominant scaling exponents cluster around approximately $0.28$ for the healthy condition and $0.22$ for the faulty condition. Furthermore, the healthy gearbox exhibits longer and wider spectral tails than the faulty one, indicating richer variability in scaling behavior. These distinctions in fractal space enable clear characterization of faulty machines based on their vibration signatures.

To assess robustness, we plot the multifractal features of all healthy and faulty gearbox datasets in Fig.~\ref{fig4}. The fractal features of machines belonging to the same category (healthy or faulty) are closely grouped and exhibit consistent structure. Conversely, the separation between the two categories is substantial, supporting reliable automated classification of gearbox health using the proposed multifractal analysis framework.

\begin{figure*}[t]
	\centering
	\begin{subfigure}{.265\textwidth}
		\includegraphics[width=\linewidth]{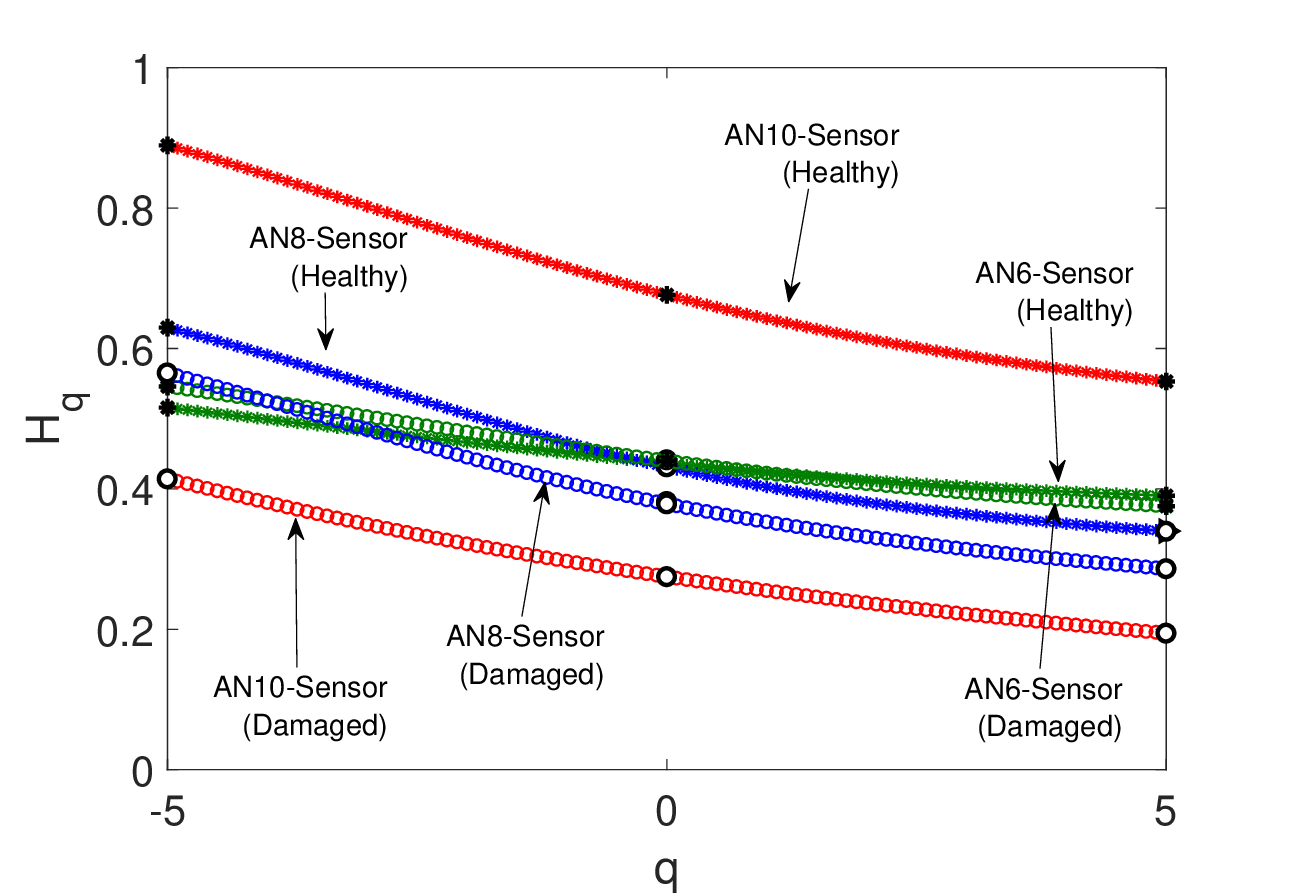}
	\end{subfigure}
	\hspace{-4mm}
	\begin{subfigure}{.265\textwidth}
		\includegraphics[width=\linewidth]{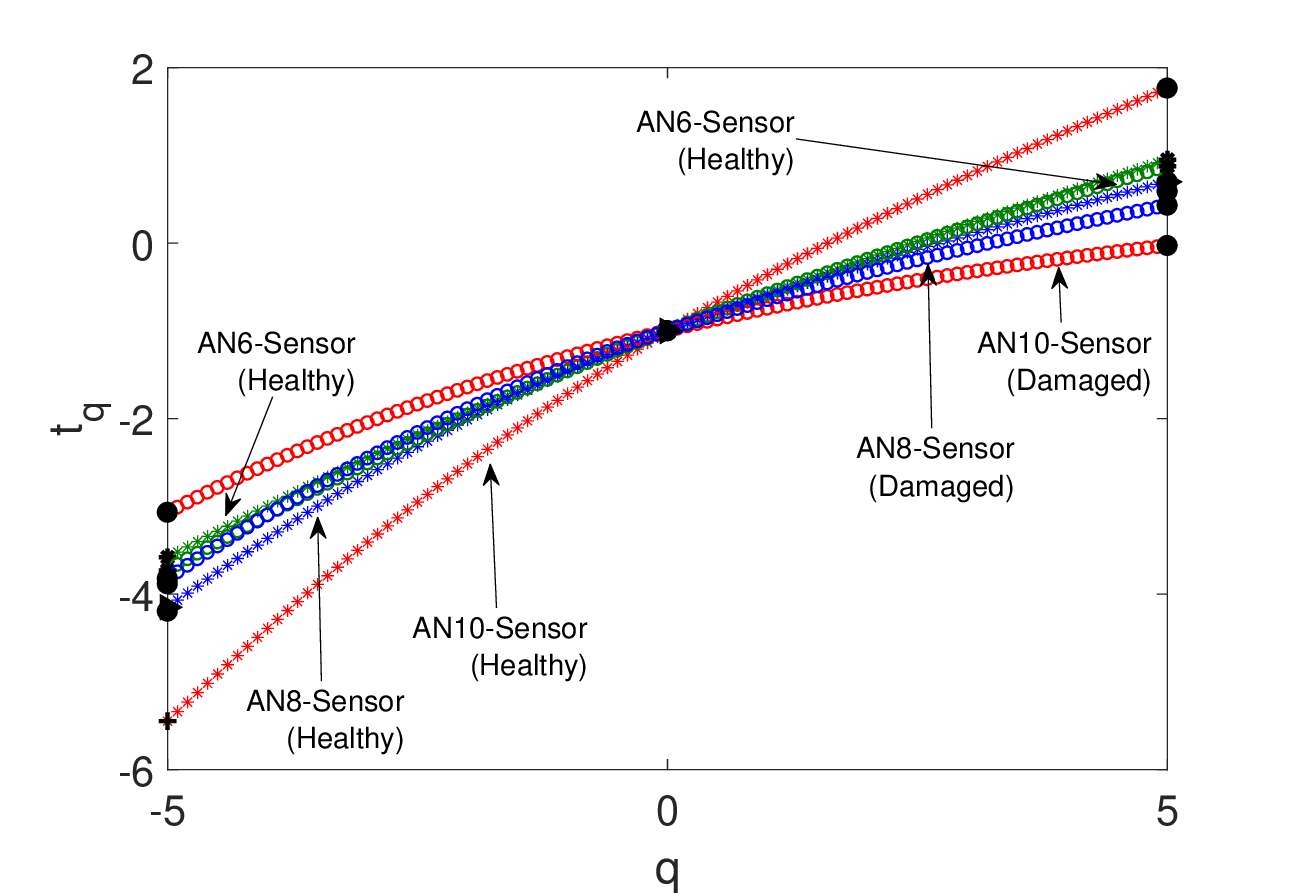}
	\end{subfigure}
	\hspace{-4mm}
	\begin{subfigure}{.265\textwidth}
		\includegraphics[width=\linewidth]{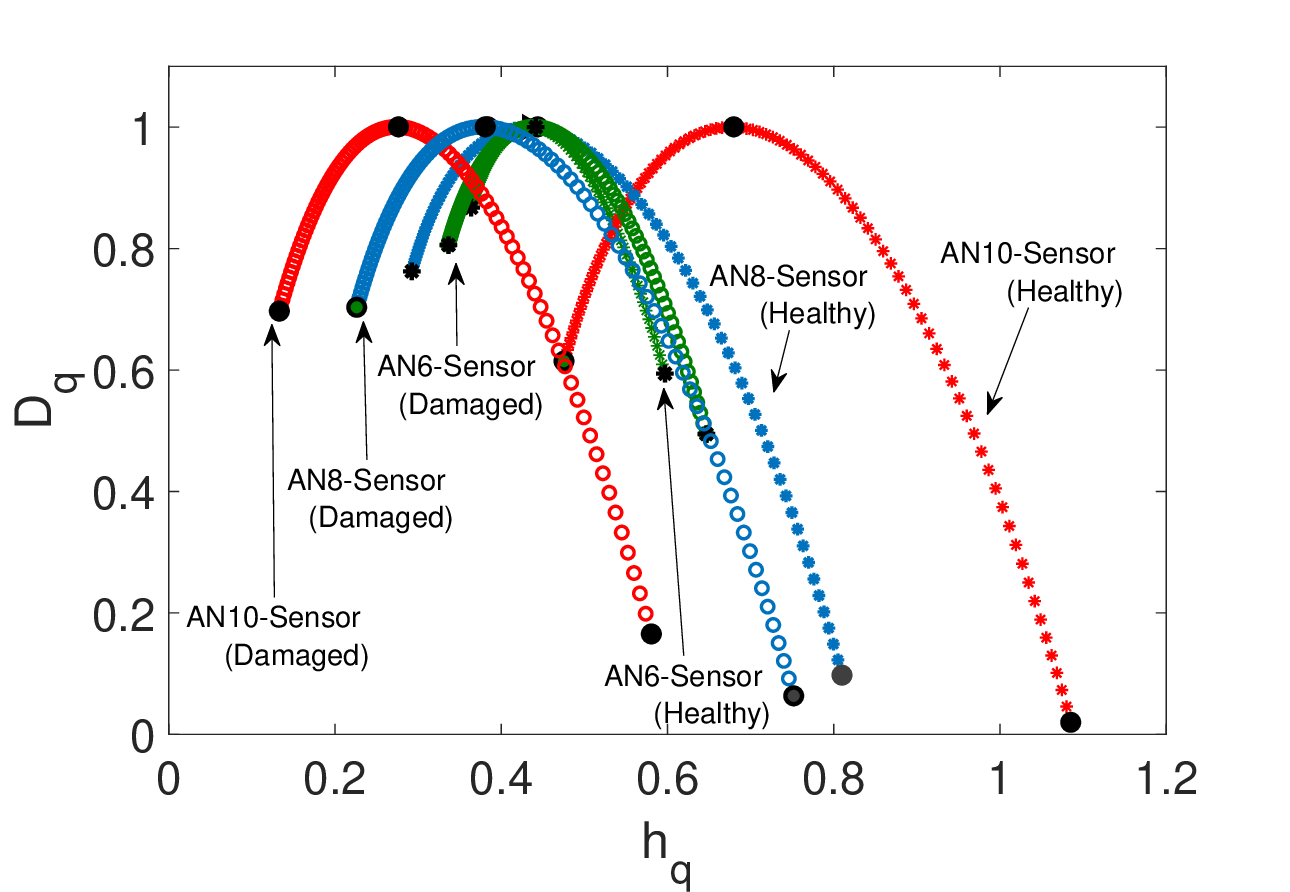}
	\end{subfigure}
	
	\begin{subfigure}{.265\textwidth}
		\includegraphics[width=\linewidth]{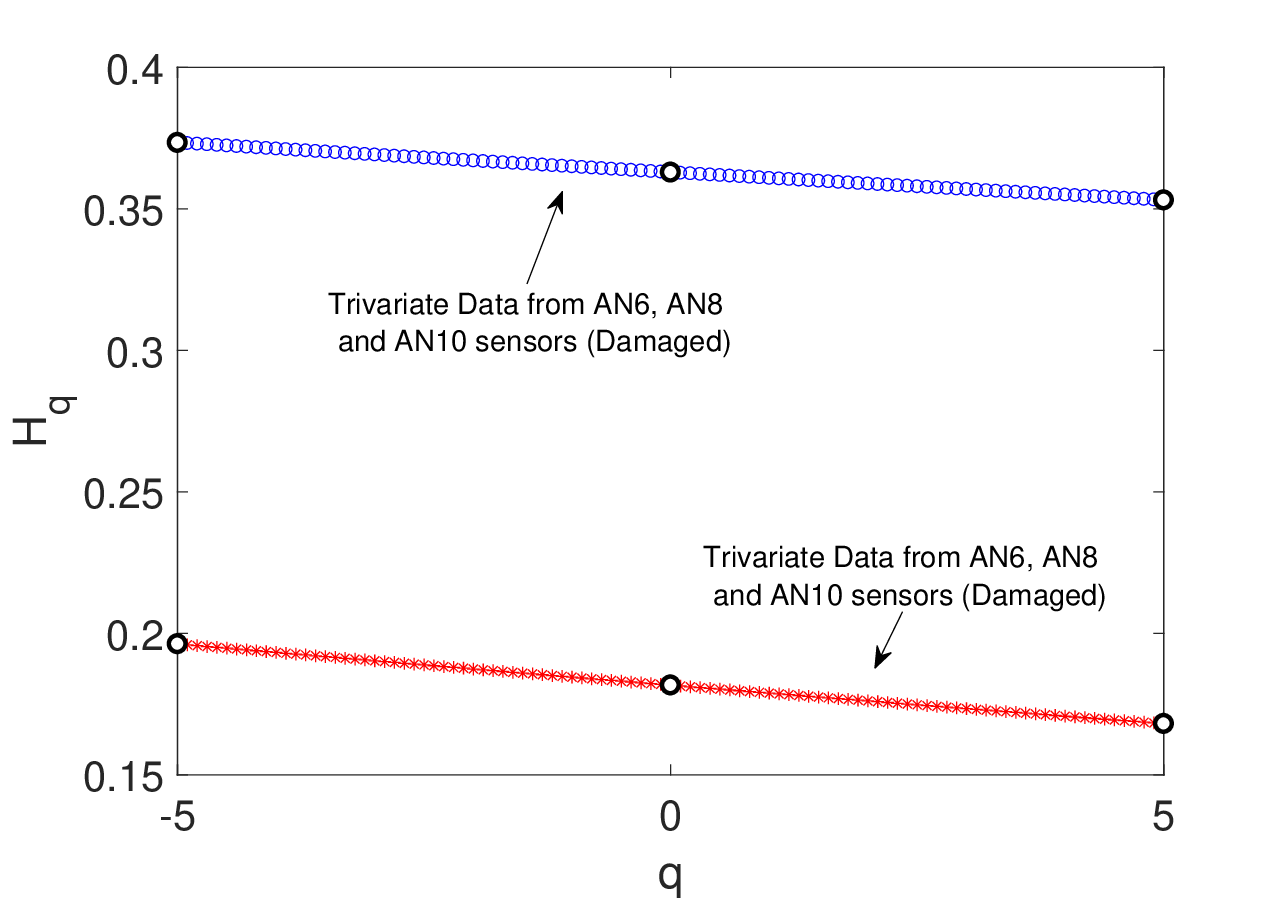}
		\caption{}
				\label{fig5a}
	\end{subfigure}
	\hspace{-4mm}
	\begin{subfigure}{.265\textwidth}
		\includegraphics[width=\linewidth]{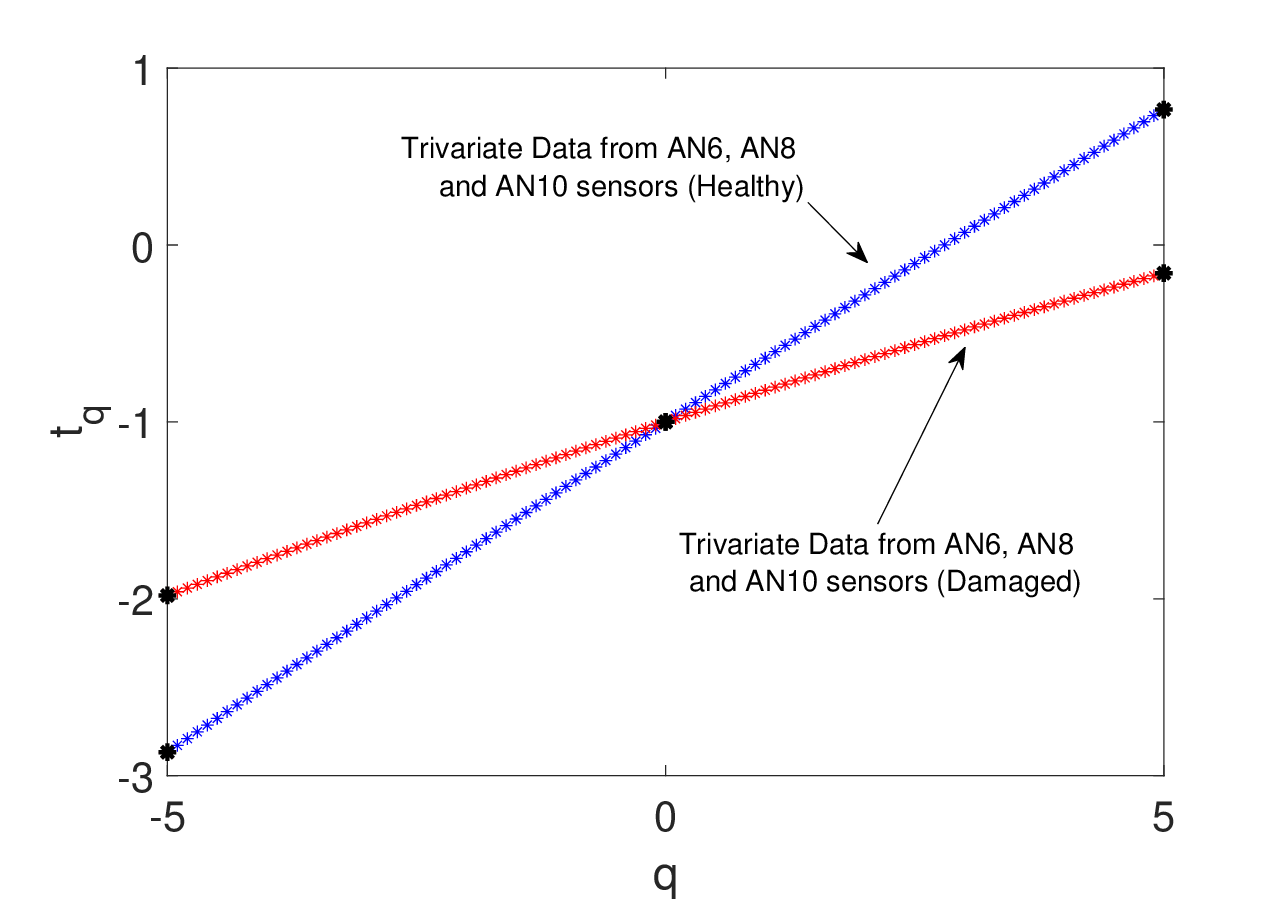}
		\caption{}
				\label{fig5b}
	\end{subfigure}
	\hspace{-4mm}
	\begin{subfigure}{.265\textwidth}
		\includegraphics[width=\linewidth]{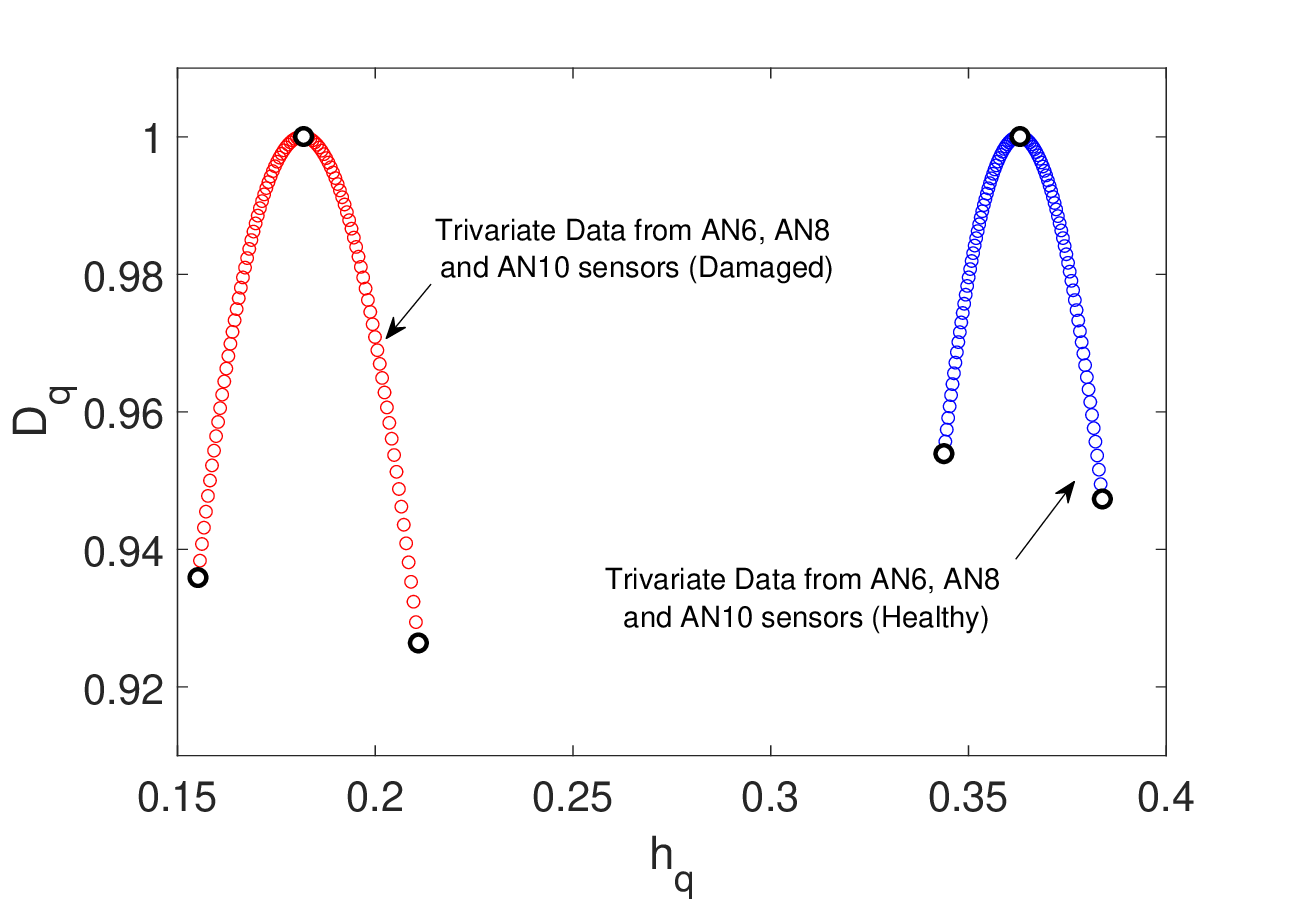}
		\caption{}
				\label{fig5c}
	\end{subfigure}
	\caption{Comparison of multifractal features using 1D DFA and the proposed GMDFA for AN6, AN8 and AN8 channels of Healthy1 and Damaged1 datasets.}
	\label{fig5}
	\vspace{-5mm}
\end{figure*}

Finally, we compare the proposed FM-MFDFA method with the conventional 1D MFDFA approach for wind turbine gearbox fault diagnosis, as shown in Fig.~\ref{fig5}. The top row of Fig.~\ref{fig5} shows the fractal features of individual channels AN6, AN8, and AN10 for both healthy and faulty gearboxes using the 1D approach. Channels AN6 (green) and AN8 (blue) exhibit nearly overlapping fractal features under both conditions, providing no conclusive evidence of faults. In contrast, channel AN10 (red) shows noticeable differences between healthy and faulty states. This suggests that relying solely on 1D multifractal analysis of individual channels may fail to detect faults if fault-related features are unevenly distributed across sensors.

By comparison, the multichannel fractal analysis of the trivariate signal (AN6, AN8, AN10), obtained using the proposed FM-MFDFA method (bottom row of Fig.~\ref{fig5}), reveals clear and unambiguous fault signatures. This highlights the importance of multichannel analysis: even when individual channels do not exhibit strong fault indicators, their joint behavior across channels provides reliable evidence of gearbox degradation. Therefore, multichannel FM-MFDFA is essential for robust fault diagnosis in multisensor vibration datasets.

\section{discussion}
The key contributions of this study can be summarized as follows: (i) a matrix norm based on the Mahalanobis distance that quantifies fluctuations by weighting channel contributions according to their statistical covariance;  
(ii) the use of this norm to construct a fully multivariate fluctuation function that enables multifractal analysis of correlated multichannel signals; and  (iii) a multichannel fault-diagnosis framework based on the proposed FM-MFDFA method, capable of capturing complex, cross-sensor fault signatures that remain undetected in conventional one-dimensional MFDFA approaches.

The experimental evaluation on wind turbine gearbox vibration data highlights the strong discriminative capability of the proposed approach. The FM-MFDFA method effectively captures structural differences between the time-series data of healthy and faulty machines, particularly through the separation observed in their fractal spectra and associated multifractal features. The inclusion of the Mahalanobis distance ensures that the fluctuation function respects the geometry of the multichannel data space, effectively normalizing for variance differences and capturing latent inter-dependencies. This way, the cross-channel correlations of multivariate data, often overlooked in traditional analyses, are included with fault analysis thereby playing an integral role in characterizing the underlying dynamics.

An important component of the framework is the use of MVMD to decompose the vibration data into intrinsic mode functions (IMFs). By examining the multifractal features of individual IMFs for both healthy and faulty conditions, we demonstrated that the first few IMFs consistently carry the most fault-relevant information. This aligns with the physical interpretation of MVMD, where low-order IMFs typically represent structured, fault-related oscillatory modes, while higher IMFs tend to be dominated by high-frequency noise. The proposed approach therefore enables a more targeted reconstruction of fault-relevant vibration components prior to multi-fractal analysis.

The comparison between standard one-dimensional MFDFA and the proposed FM-MFDFA method further underscores the advantages of a multivariate formulation. Results show that univariate fractal features from individual channels may fail to reveal clear fault signatures when those features are unevenly distributed or present with varying intensity across sensors. In contrast, the FM-MFDFA method integrates information across all channels through a covariance-aware matrix norm, yielding unambiguous detection of fault signatures. This reinforces the necessity of multichannel analysis in vibration-based condition monitoring, particularly in systems where sensor responses are strongly coupled through the underlying mechanical structure. 

\section{Conclusions and Discussion}
This work introduced a fully multivariate generalization of the multifractal detrended fluctuation analysis (MFDFA) framework by developing a novel matrix norm based on the Mahalanobis distance. This matrix norm provides a principled way to incorporate cross-channel covariance into the fluctuation measurement process, thereby overcoming the inherent limitations of existing univariate and partially multivariate MFDFA formulations. Building on this theoretical contribution, we proposed a complete fault-diagnosis framework for multichannel vibration data that explicitly accounts for the inter-channel dependencies-an essential characteristic of real-world sensor systems such as wind turbine gearboxes. Future work may extend this framework to real-time monitoring applications by reducing computational complexity or employing approximations of the covariance matrix. Moreover, coupling the proposed FM-MFDFA features with advanced classifiers or deep learning models may further improve the accuracy and generalizability of fault-diagnosis systems in large-scale industrial environments.

\section{Appendixes}
\subsection{Mahalanobis distance Matrix norm}
We know that inverse of a symmetric and positive definite matrix, $\Sigma=\Sigma^T \succ 0$, is also symmetric and positive definite matrix $\Sigma^{-1}= {\Sigma^{-1}}^T\succ 0$; hence, the only way 
\begin{equation}
\begin{split}
&\|\pmb{Z}\|_{pq^\Sigma} = \|\ \{\pmb{z}_i\}_{u=1}^U\}_{v=1}^V \ \|_{pq^\Sigma} =0, \\
&\quad \mathit{iff} \ \pmb{z}_{u,v}=\pmb{0} \ \ \forall \ u \ \& \ v, i.e., \ \pmb{Z} = \pmb{0},
\end{split}
\end{equation}
	
Also, since $\|\pmb{Z}\|_{pq^\Sigma}$ is based on a quadratic function $\sqrt{\pmb{z}^T\Sigma^{-1}\pmb{z}}$ with a positive definite matrix $\Sigma^{-1}\succ 0$, 
\begin{equation}
\|\ \{\pmb{z}_{u,v}\}_{u=1}^U\}_{v=1}^V \ \|_{pq^\Sigma} >0, \ \forall \ \ \pmb{z}_{u,v}\ne\pmb{0}.
\end{equation}

Moreover, it is also trivial to show that $\|\mathcal{K}\pmb{Z}\|_{pq^\Sigma} = |\mathcal{K}|\cdot \|\pmb{Z}\|_{pq^\Sigma}$ for a scalar $\mathcal{K}$, as follows
	\begin{equation}
		\small
		\begin{split}
		\|\mathcal{K}\pmb{Z} \|_{pq^\Sigma} &= \left (\sum_{u=1}^{U} \Big(\ \sum_{v=1}^{V}\Big(\sqrt{(\mathcal{K}\pmb{z}_{u,v}^T)\ \Sigma^{-1} \ (\mathcal{K}\pmb{z}_{u,v})}\ \Big)^p\ \Big)^\frac{q}{p}\right)^{\frac{1}{q}}\\
		&= \left (\sum_{u=1}^{U} \Big(\ \sum_{v=1}^{V}\Big(\mathcal{K}\sqrt{(\pmb{z}_{u,v}^T)\ \Sigma^{-1} \ (\pmb{z}_{u,v})}\ \Big)^p\ \Big)^\frac{q}{p}\right)^{\frac{1}{q}}\\
		&= \mathcal{K}\left (\sum_{u=1}^{U} \Big(\ \sum_{v=1}^{V}\Big(\sqrt{(\pmb{z}_{u,v}^T)\ \Sigma^{-1} \ (\pmb{z}_{u,v})}\ \Big)^p\ \Big)^\frac{q}{p}\right)^{\frac{1}{q}}\\
		&= \mathcal{K}\|\pmb{Z} \|_{pq^\Sigma}
		\end{split}
	\end{equation}
	To prove the triangular inequality, we perform spectral decomposition of symmetric matrices that results in a diagonal $M\times M$ matrix $\Lambda=diag(\lambda_1,\lambda_2,\ldots,\lambda_n)$ and an orthogonal  $M\times M$ matrix $Q$ (i.e., $Q^TQ=I_{M\times M}$ where $I_{M\times M}$ denotes an identity matrix), such that $Q^T=Q^{-1}$ and
	\begin{equation}\label{A1}
		\Sigma=Q^T\Lambda Q.
		\vspace{-1mm}
	\end{equation}
	
	By definition, $\Sigma$ is positive-definite matrix that means 
		$\lambda_1>0,\lambda_2>0,\ldots,\lambda_m>0$, hence a matrix $S$ may be defined
	\begin{equation} \label{A2}
		S = \Lambda^{\frac{1}{2}} Q = diag(\sqrt{\lambda_1},\sqrt{\lambda_2},\ldots,\sqrt{\lambda_n})\ Q,
	\end{equation} 
	\textit{such that}
	\begin{equation}\label{A3}
		\Sigma= S^T S.
	\end{equation}
	
	By using \eqref{A2} in \eqref{Eq01}, we get
	\begin{equation}\label{A4}
		\small
		\begin{split}
				\|\pmb{Z} \|_{pq^\Sigma} &= \left (\sum_{u=1}^{U} \Big(\ \sum_{v=1}^{V}\Big(\sqrt{(\pmb{z}_{u,v}^T)\ S^T \ S \ (\pmb{z}_{u,v})}\ \Big)^p\ \Big)^\frac{q}{p}\right)^{\frac{1}{q}}\\
				&= \left (\sum_{u=1}^{U} \Big(\ \sum_{v=1}^{V}\Big(\sqrt{(S\pmb{z}_{u,v})^T\ (S\pmb{z}_{u,v})}\ \Big)^p\ \Big)^\frac{q}{p}\right)^{\frac{1}{q}}
		\end{split}
	\end{equation}

	Now, let us set $\tilde{\pmb{z}}_i=S\pmb{z}_i$ then
	\begin{equation}\label{A5}
		\small
		\begin{split}
		\|\pmb{Z}\|_{pq^\Sigma}&= \left (\sum_{u=1}^{U} \Big(\ \sum_{v=1}^{V}\Big(\sqrt{(\tilde{\pmb{z}}_{u,v})^T\ (\tilde{\pmb{z}}_{u,v})}\ \Big)^p\ \Big)^\frac{q}{p}\right)^{\frac{1}{q}}\\
		&=\|\tilde{\pmb{Z}}\|_{pq}
		\end{split}
\end{equation}
where $\tilde{\pmb{Z}}=\tilde{\pmb{z}}_{u,v}\}_{u=1}^U\}_{v=1}^V$.

Following from \eqref{A5}, it is straight forward to show that
	\begin{equation}\label{A6}
		\small
		\begin{split}
			\|&\pmb{Z}_1+\pmb{Z}_2\|_{pq^\Sigma}\\ & = \left (\sum_{u=1}^{U} \Big(\sum_{v=1}^{V}\Big( \sqrt{({\pmb{z}_1}_{u,v}+{\pmb{z}_2}_{u,v})^T \Sigma^{-1}  ({\pmb{z}_1}_{u,v}+{\pmb{z}_2}_{u,v})}\ \Big)^p \Big)^\frac{q}{p}\right)^{\frac{1}{q}}\\
	        & = \left (\sum_{u=1}^{U} \Big(\sum_{v=1}^{V}\Big( \sqrt{({\pmb{z}_1}_{u,v}+{\pmb{z}_2}_{u,v})^T S^T\ S \ ({\pmb{z}_1}_{u,v}+{\pmb{z}_2}_{u,v})}\ \Big)^p \Big)^\frac{q}{p}\right)^{\frac{1}{q}}\\
			& = \left (\sum_{u=1}^{U} \Big(\sum_{v=1}^{V}\Big( \sqrt{(S{\pmb{z}_1}_{u,v}+S{\pmb{z}_2}_{u,v})^T (S{\pmb{z}_1}_{u,v}+S{\pmb{z}_2}_{u,v})}\ \Big)^p \Big)^\frac{q}{p}\right)^{\frac{1}{q}}\\
			& = \left (\sum_{u=1}^{U} \Big(\sum_{v=1}^{V}\Big( \sqrt{(\tilde{\pmb{z}}_{1_{u,v}}+\tilde{\pmb{z}}_{2_{u,v}})^T (\tilde{\pmb{z}}_{1_{u,v}}+\tilde{\pmb{z}}_{2_{u,v}})}\ \Big)^p \Big)^\frac{q}{p}\right)^{\frac{1}{q}}\\
			&=\|\tilde{\pmb{Z}}_1+\tilde{\pmb{Z}}_2\|_{pq}
		\end{split}
	\end{equation}
	Since, we already know that $L_{pq}$ matrix norm fulfills following inequalities \cite{}, which means, 
	\begin{equation}\label{A7}
		\begin{split}
		&\|\tilde{\pmb{Z}}_1+\tilde{\pmb{Z}}_2\|_2\ \leq\ \|\tilde{\pmb{Z}}_1\|_2+\|\tilde{\pmb{Z}}_2\|_2,\\
		&\|\tilde{\pmb{Z}}_1 \cdot \tilde{\pmb{Z}}_2\|_2\ \leq\ \|\tilde{\pmb{Z}}_1\|_2 \cdot \|\tilde{\pmb{Z}}_2\|_2,
		\end{split}
	\end{equation}
	that essentially means
	\begin{equation}\label{A8}
		\begin{split}
		&\|\pmb{Z}_1+\pmb{Z}_2\|_{pq^\Sigma} \leq \|\pmb{Z}_1\|_{pq^\Sigma}+\|\pmb{z}_2\|_{pq^\Sigma}\\
		&\|\pmb{Z}_1\cdot \pmb{Z}_2\|_{pq^\Sigma} \leq \|\pmb{Z}_1\|_{pq^\Sigma} \cdot \|\pmb{z}_2\|_{pq^\Sigma}
		\end{split}
	\end{equation}
\subsection{Interpretation of \texorpdfstring{$L_{pq^\Sigma}$}{LpqSigma} norm}
$L_{pq^\Sigma}$ norm, introduced in proposition \ref{definition3}, that computes norm of multivariate data by incorporating its cross channel dependencies using Mahalanobis distance, see \eqref{Eq11}, is a multichannel generalization of the $L_{pq^E}$ norm \eqref{Eq05}. Here, it is easy to show that $L_{pq^E}$ norm is a special case of the $L_{pq^\Sigma}$ norm when $\tiny{\Sigma=I_{m\times m}}$, as follows
\begin{equation}\label{B1}
\begin{split}
\small{\|\pmb{Z}\|_{pq^{\Sigma=I_{m\times m}}} =\|\pmb{Z}\|_{pq^E}.}
 \end{split}
\end{equation}
	
Secondly, when $\Sigma=\boldsymbol{\sigma}^T I_{m\times m}$ is a diagonal matrix where the vector $\boldsymbol{\sigma} = [\sigma_1,\sigma_2,\ldots,\sigma_m]^T$ contains channel variances, $\|\pmb{Z}\|_{pq^\Sigma}$ is given by
	
\begin{equation}\label{B2}
\small
\begin{split}
		&\small{\|\pmb{Z}\|_{pq^{\Sigma=\boldsymbol{\sigma}^T I_{M\times M}}}=}\\
		&\left (\sum_{u=1}^{U} \left(\ \sum_{v=1}^{V}\left(\sqrt{\frac{z_{{u,v}_1}^2}{\sigma_1^2} + \ldots + \frac{z_{{u,v}_M}^2}{\sigma_M^2}}\ \right)^p\ \right)^\frac{q}{p}\right)^{\frac{1}{q}} = \|\overline{\pmb{Z}}\|_{pq^E}
		\vspace{-1mm}
\end{split}
\end{equation}
\textit{where $\overline{\pmb{Z}}=\overline{\pmb{z}}_{u,v}\}_{u=1}^U\}_{v=1}^V$ and $\overline{\pmb{z}}=[\frac{z_{1}}{\sigma_1},\ldots,\frac{z_{m}}{\sigma_M}]^T$.}
	
Finally, in case of correlated multivariate data, Mahalanobis norm essentially computes $L_2$ norm by un-correlating the variance normalized vector observations as depicted in Fig. \ref{fig01}. Its easy to show this for a special case of bivariate data, having covariance matrix $\Sigma = \tiny{ 
		\begin{pmatrix}
			\sigma_1^2 & \rho \sigma_1 \sigma_2 \\
			\rho \sigma_1 \sigma_2 & \sigma_2^2
	\end{pmatrix}}$ where $\rho$ is the correlation coefficient. $\|\pmb{Z}\|_{pq^\Sigma}$ can be rewritten as
	\begin{equation}\label{B3}
	\small
	\begin{split}
			&\small{\|\pmb{Z}\|_{pq^{\Sigma=\boldsymbol{\sigma}^T I_{m\times m}}}=}\\
			&\small{\left (\sum_{u=1}^{U} \left(\ \sum_{v=1}^{V}\left(\frac{1}{1-\rho^2}\ldots\Big(
			\frac{z_{1_{u,v}}^2}{\sigma_1^2} + \frac{z_{2_{u,v}}^2}{\sigma_2^2}\ - \frac{2\rho z_{1_{u,v}} z_{2_{u,v}}}{\sigma_1 \sigma_2}\Big) \right)^\frac{p}{2}\ \right)^\frac{q}{p}\right)^{\frac{1}{q}} }
			\end{split}
	\end{equation}
Clearly, MD is subtracting the variance normalized covariance factors $\frac{2\rho z_{1_{u,v}} z_{2_{u,v}}}{\sigma_1 \sigma_2}$ from before computing the norm $\frac{z_{{u,v}_1}^2}{\sigma_1^2} + \frac{z_{{u,v}_2}^2}{\sigma_2^2}$ which is essentially a whitening operation, i.e., removal of cross channel correlations before computing the Euclidean norm. That means proposed multichannel norm function based on MD performs uncorrelation operation on the data before computing its norm because ED is blind to cross channel correlations.

\bibliographystyle{IEEEtran}
\bibliography{References}

\end{document}